\documentclass[11pt]{article}
\usepackage{jheppubmod}
\pdfoutput=1
\usepackage{psfrag}
\usepackage{array}
\usepackage{amssymb}
\usepackage{amsmath}
\usepackage{amsthm}
\usepackage{mathtools}
\usepackage{graphicx}
\usepackage{placeins}
\usepackage[labelsep=quad]{subcaption}

\usepackage[square,sort&compress,numbers]{natbib}
\usepackage{epsfig}
\def\lmodep{v^c_{\omega,\ell}}
\def\rmodep{u^c_{\omega,\ell}}
\def\lmodem{v^+_{\omega,\ell}}
\def\rmodem{u^+_{\omega,\ell}}

\def\rpinmode{\phi^{in,+}}
\def\rcoutmode{\phi^{out,c}}
\def\normalization{{\cal N}}

\def\lmodeinsidep{\tilde{v}^+_{\omega,\ell}}
\def\rmodeinsidep{\tilde{u}^+_{\omega,\ell}}
\def\lmodeinsidem{\tilde{v}^-_{\omega,\ell}}
\def\rmodeinsidem{\tilde{u}^-_{\omega,\ell}}

\def\boga{{\cal A}_{\omega,\ell}}
\def\bogb{{\cal B}_{\omega,\ell}}
\def\bogas{\boga^*}
\def\bogbs{\bogb^*}
\def\bogainside{\tilde{\cal A}_{\omega,\ell}}
\def\bogbinside{\tilde{\cal B}_{\omega,\ell}}
\def\bogasinside{\bogainside^*}
\def\bogbsinside{\bogbinside^*}
\def\bogal{{\cal A}}

\def\tune{{\cal T}}
\def \sharp{\mathfrak{s}}
\def\anha{{\frak a}}
\def\anhb{{\frak b}}
\def\anhc{{\frak c}}
\def\anhd{{\frak d}}
\def\anhe{{\frak e}}
\def\anhf{{\frak f}}

\def\anhet{\tilde{\frak e}}
\def\anhft{\tilde{\frak f}}
\def \sph{{\Omega}}

\title{Quantum aspects of charged black holes in de-Sitter space}
\author[a]{Pushkal Shrivastava}
\affiliation[a]{Centre for High Energy Physics, Indian Institute of Science, CV Raman Avenue, Bengaluru - 560012, India}
\emailAdd{pushkals@iisc.ac.in}
\date{}

\abstract{It has been argued that generic classical perturbations to asymptotically de-Sitter Reissner-Nordtsr\"om (RN-dS) black holes may violate strong cosmic censorship conjecture. In this paper, we analyze whether quantum corrections can restore the conjecture. We study a quantum scalar field in RN-dS geometry and analyze the smoothness of a state across various horizons using the criteria developed in arXiv:1910.02992. Since de-Sitter black holes have a cosmological horizon, that typically radiates at a different temperature than the event horizon, the existence of a quantum state which is regular everywhere in the exterior region is non-trivial. We find such states for spherically symmetric black holes in arbitrary dimensions. We then demonstrate that such states are singular at the inner horizon of RN-dS black holes in various dimensions. Hence, quantum fluctuations are sufficient to restore the strong cosmic censorship conjecture in RN-dS.}

\begin{document}
\maketitle

\section{Introduction \label{s_intro}}
The question of whether or not classical general relativity is deterministic has garnered the attention of researchers for decades. Black hole solutions such as Reissner-Nordstr\"om and Kerr seem to provide examples where determinism fails. Apart from the event horizon, these spacetimes have an inner horizon as well, which is also a Cauchy horizon. Initial data on a Cauchy slice cannot determine the evolution of fields beyond the Cauchy horizon. However, the spacetime can be extended beyond the inner horizon as curvature invariants do not diverge, and there are causal trajectories which approach the inner horizon in finite proper time. 

Before concluding that determinism fails in gravity, it is necessary to understand the effects of perturbations. If generic perturbations destabilize the inner horizon, then determinism would be restored. Spacetimes with an inner horizon provide a testing ground for the validity of strong cosmic censorship conjecture \cite{penrose1968structure}, which contends that the Cauchy problem in general relativity is well-posed for generic initial data. Hence, loss of determinism should occur only in very special spacetimes, which are irrelevant for physical discussions.

The effects of classical perturbations on the stability of inner horizon has been studied extensively \cite{Simpson:1973ua,mcnamara1978instability,chandrasekhar1982crossing,Poisson:1989zz,Poisson:1990eh,Ori:1991zz,Ori:1992zz,Mellor:1989ac,Chambers:1994ap,Brady:1996za,Dafermos:2003wr,Dafermos:2002ka, Murata:2013daa,christodoulou2012formation,Bhattacharjee:2016zof,Hod:2018dpx,Mo:2018nnu,Hod:2018lmi,Luk:2017jxq,Luna:2018jfk,Dias:2018ynt,Dafermos:2012np,Dafermos:2017dbw,dafermos2018rough,Cardoso:2017soq}. Interestingly, it has been argued that Reissner-Nordstr\"om black holes in de-Sitter spacetimes violate strong cosmic censorship conjecture close to extremality \cite{Cardoso:2017soq, Dias:2018ufh}. 

On the other hand, the effects of quantum fluctuations on the stability of the inner horizon is relatively unexplored, and has gained traction only recently \cite{Papadodimas:2019msp,Sela:2018xko, Dias:2019ery,Lanir:2018vgb,Zilberman:2019buh,Balasubramanian:2019qwk,Emparan:2020rnp,Hartnoll:2020rwq,Emparan:2020znc,Hollands:2019whz,Hollands:2020qpe}. One common approach is to determine whether or not the renormalized quantum stress tensor diverges at the inner horizon. However, due to the complexity of the computation and ambiguities in the renormalization procedure, this may not always be the most efficient approach. Therefore, we need simpler tests which could shed some light on this problem. 

To this end, a simple criteria for testing smoothness of a quantum state across a null surface in arbitrary spacetime was recently developed in \cite{Papadodimas:2019msp}. The authors of \cite{Papadodimas:2019msp} studied a scalar field propagating in a fixed geometry. They demonstrated that a quantum state is smooth across a null surface only if modes defined by integrating the field in local Rindler coordinates near the null surface are entangled in a specific way. To reach this conclusion, they assumed that the two-point correlation function of field insertions reduces to the two-point function in flat spacetime if the insertions are taken close to each other. Since the contribution of this term is subtracted to get a finite renormalized stress tensor, this assumption is a necessity for a non-divergent quantum stress tensor. Using this, the authors were able to develop a simple test for strong cosmic censorship in asymptotically anti-de-Sitter spacetimes and rule out violations in Reissner-Nordstr\"om black holes in AdS. 

In this work, we use this criteria for smoothness to study quantum aspects of eternal charged black holes in de-Sitter space. Since black holes in de-Sitter space are additionally endowed with a cosmological horizon, it is interesting to understand whether there exists a quantum state which is regular everywhere in the exterior, i.e., the expectation value of the stress tensor is finite everywhere in the region between outer horizon and cosmological horizon. This is an interesting question even for Schwarzschild black holes. Since the cosmological horizon and event horizon radiate at different temperatures, it is often expected that no such quantum state should exist \cite{Hiscock:1989yw}. This issue has received some attention for two-dimensional black holes  \cite{Markovic:1991ua,Tadaki:1990cg,Ghafarnejad:2006pm}, where it has been argued that a smooth state should exist. However, to the best of our knowledge, status of higher dimensional black holes remain unclear. Regularity of a quantum state in a charged black holes in four dimensions, was analyzed in \cite{Hollands:2019whz}. The existence of Hadamard states in the exterior was argued. In this paper, we explicitly construct quantum states which satisfy the smoothness criteria in \cite{Papadodimas:2019msp}, everywhere in the exterior of any spherically symmetric non-extremal eternal black hole, in \emph{any} dimensions. This strongly suggests the possibility of existence of quantum states that are regular everywhere in the exterior. 

We then show that such states always lead to instability at the inner horizon. Hence, quantum fluctuations restore the strong cosmic censorship. Our results are in agreement with recent works \cite{Hollands:2019whz,Hollands:2020qpe}. 

The paper is structured as follows. We describe the classical geometry and define various coordinate patches of RN-dS in section \ref{s_geometry}. In section \ref{s_exterior} we determine the quantum state that is regular everywhere in the exterior. In section \ref{s_interior} and \ref{s_numerics} we show that the quantum state is singular at the inner horizon. Finally, we conclude the paper in section \ref{s_conclusions}.



\section{Geometry of Reissner-Nordstr\"om-de-Sitter black holes\label{s_geometry}}
We begin by briefly reviewing the geometry of asymptotically de-Sitter Reissner-Nordstrom black holes. The metric of such a black hole in $d+1$ spacetime dimensions is given by,
\begin{equation}\label{metric}
\begin{split}
ds_{d+1}^2 &= - f(r) dt^2 + {dr^2 \over f(r)} + r^2 d\Omega_{d-1}^2, \qquad
f(r) = 1 -  r^2 - {A \over r^{d-2}} + {B^2 \over r^{2(d-2)}}.
\end{split}
\end{equation}
We have set the de-Sitter scale to unity. The parameters $A$ and $B$ are related to mass and charge of the black hole. Roots of $f(r)$ determine the location of the horizons. Physically allowed set of parameters admit three real and positive roots. The largest root corresponds to the location of the cosmological horizon, which we denote by $r_c$. Second largest root, $r_+$, is location of the outer horizon. While, the smallest positive root, $r_-$, is location of the inner horizon. 

The coordinates in \eqref{metric} are singular at the horizons. To define coordinate patches that are regular in the vicinity of various horizons, we first define the tortoise coordinate,
\begin{equation}
r_* = \int { dr \over f(r)}.
\end{equation}
As long as $f(r)$ has three distinct positive roots, the tortoise coordinate has the following near horizon limits.
\begin{equation}\label{tortoise}
\begin{split}
\lim\limits_{r\rightarrow r_c} r_*  &= -{\log|r-r_c| \over 2\kappa_c} + {\log({\zeta^2_c\over2\kappa_c})\over 2\kappa_c}, \\
\lim\limits_{r\rightarrow r_+} r_*  &= {\log|r-r_+| \over 2\kappa_+} - {\log({\zeta^2_+\over2\kappa_+})\over 2\kappa_+}, \\
\lim\limits_{r\rightarrow r_-} r_*  &= -{\log|r-r_-| \over 2\kappa_-} + {\log({\zeta^2_-\over2\kappa_-})\over 2\kappa_-},
\end{split}
\end{equation}
where, $\zeta$'s are related to the constant of integration. We have introduced the surface gravity at horizons,
\begin{equation}
\kappa_i = {1\over2} |f'(r_i)| , \quad i\in\{-,+,c\}.
\end{equation}
The form of the constant piece in \eqref{tortoise} is chosen so that the metric takes particularly simple form near horizon in the Kruskal coordinates. Kruskal coordinates allowing extension of spacetime beyond the outer horizon are, 
\begin{equation}\label{kruksalp}
\begin{split}
U_+ = -{\zeta_+ \over \kappa_+} e^{-\kappa_+(t-r_*)}, \qquad 	V_+ = {\zeta_+ \over \kappa_+} e^{\kappa_+(t+r_*)}  , \qquad r>r_+\\
U_+ = {\zeta_+ \over \kappa_+} e^{-\kappa_+(t-r_*)}, \qquad 	V_+ = {\zeta_+ \over \kappa_+} e^{\kappa_+(t+r_*)}  , \qquad r<r_+.
\end{split}
\end{equation}
The coordinates above are well defined in the range $r_-<r<r_c$. In this coordinate system, all components of metric are regular near the outer horizon, $U_+V_+=0$.
\begin{equation}\label{metricp}
\lim\limits_{r\rightarrow r_+} ds_{d+1}^2 = -dU_+dV_+ +r^2 d\Omega_{d-1}^2.
\end{equation}

We introduce the second set of Kruskal coordinates that allow smooth extension beyond the inner horizon. These coordinates are well defined in the range $0<r<r_+$. 
\begin{equation}\label{kruksalm}
U_- = -{\zeta_- \over \kappa_-} e^{\kappa_-(t-r_*)}, \qquad 	V_- = -{\zeta_- \over \kappa_-} e^{-\kappa_-(t+r_*)}.
\end{equation}
The metric near the inner horizon, $U_-V_-=0$, is
\begin{equation}\label{metricn}
\lim\limits_{r\rightarrow r_-} ds_{d+1}^2 = -dU_-dV_- +r^2 d\Omega_{d-1}^2.
\end{equation}

\begin{figure}
	\centering
	\includegraphics[width=6cm]{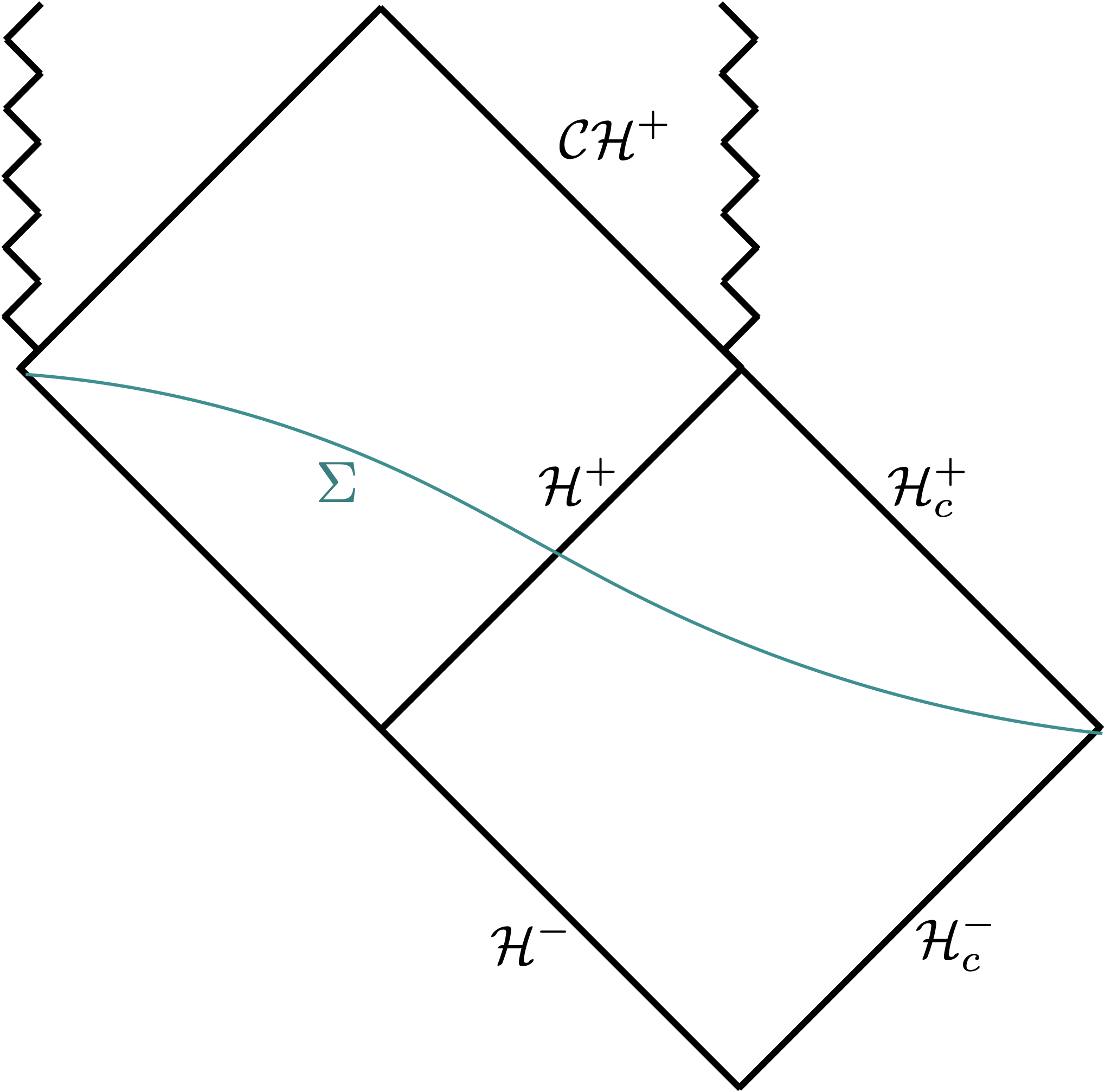}
	\caption{Penrose diagram of Reissner-Nordstr\"om black holes in de-Sitter space. \label{fig_penrosernds}}
\end{figure}

Finally, we introduce the third set of Kruskal coordinates that allow smooth extension of spacetime beyond the cosmological horizon.
\begin{equation}\label{kruksalc}
U_c = {\zeta_c \over \kappa_c} e^{\kappa_c(t-r_*)}, \qquad 	V_c = -{\zeta_c \over \kappa_c} e^{-\kappa_c(t+r_*)}.
\end{equation}
The metric near the cosmological horizon, $U_cV_c=0$, is
\begin{equation}\label{metricc}
\lim\limits_{r\rightarrow r_c} ds_{d+1}^2 = -dU_cdV_c +r^2 d\Omega_{d-1}^2.
\end{equation}
The Penrose diagram of the relevant parts of the spacetime is given in Figure \ref{fig_penrosernds}.

\FloatBarrier



\section{A quantum state regular in the exterior \label{s_exterior}}
In order to study the quantum effects, we consider a quantum field propagating in the fixed geometry of asymptotically de-Sitter Reissner-Nordstr\"om black hole \eqref{metric}. For simplicity, we consider a free massless scalar field, however, our analysis can be trivially extended to include mass, spin, and interactions. Using rotation and $t-$translation symmetry of the system, the scalar field operator can be expanded as
\begin{equation}\label{modeexp}
\phi = \sum_{\vec{\ell}\,} \int {d\omega \over\sqrt{2\pi}}{1\over \sqrt{2\omega} r^{d-1\over2} } F_{\omega,\ell}(r_*)e^{- i \omega t} Y_{\vec{\ell}}\,(\Omega) + \text{h.c.}\,.
\end{equation}
where $Y_{\vec{\ell}\,}(\Omega)$ are the spherical harmonics on $S^{d-1}$, and $\vec{\ell}\,$ collectively denotes the angular momentum quantum numbers.  
\begin{equation}
\nabla^2_{\sph} Y_{\vec{\ell}\,}(\Omega) = -\ell(\ell+d-2) Y_{\vec{\ell}\,}(\Omega),
\end{equation}
where $\nabla^2_{\sph}$ is the Laplacian on $S^{d-1}$.
The operators, $F_{\omega,\ell}$, satisfy the following radial wave equation.
\begin{equation} \label{radialwave}
\begin{split}
&	{d^2\over dr^2_*} F_{\omega,\ell} (r_*) - V_{\omega,\ell\,}(r_*)  F_{\omega,\ell}(r_*) = 0,\\
&	V_{\omega,\ell\,}(r_*) = -\omega^2 + {f(r)\over r^2} \left( \ell(\ell+d-2) +{(d-3)(d-1)\over4} f(r)+{d-1\over2} r f'(r) \right).
\end{split}
\end{equation}
The Klein-Gordon inner product on any Cauchy slice, $\Sigma$, with timelike unit normal, $n$ is given by,
\begin{equation}\label{innerkg}
(\psi_1,\psi_2) = - i \int d\Sigma\, n^{\mu}\, \psi_1 \overset{\leftrightarrow}{\partial}_{\mu}\, \psi_2^*,
\end{equation}
where $d\Sigma$ is the volume form on $\Sigma$. The future cosmological and event horizon together form a null Cauchy slice for the exterior region, $\Sigma^+ = {\cal H}^+ \cup {\cal H}_c^+$. Similarly, $\Sigma^- = {\cal H}^- \cup {\cal H}_c^-$ is also a Cauchy slice for the region between cosmological and event horizon. To define the inner product on null Cauchy slices, we can deform the metric slightly, so that these slices become spacelike and then take the deformation to zero. The inner product defined this way is well behaved and independent of the choice of deformation.
\begin{equation}\label{innernull}
(\psi_1,\psi_2) = - i \int_{{\cal H}_c^{\pm}} du\, d\Omega \, r^{d-1}\, \psi_1 \overset{\leftrightarrow}{\partial}_{u}\, \psi_2^* - i \int_{{\cal H}^{\pm}} dv\, d\Omega \, r^{d-1}\, \psi_1 \overset{\leftrightarrow}{\partial}_{v}\, \psi_2^*.
\end{equation}

\subsection{Mode expansion in the exterior}\label{s_qscalar}
We are interested in the mode expansion of the scalar field near various horizons. As we approach any of the three horizons, the potential simplifies significantly, $V_{\omega,\ell}\rightarrow-\omega^2$. Hence the scalar field near a horizon can be expanded in a basis of plane wave solutions. We define the following radial modes, in the exterior, by fixing the behavior at either cosmological or outer horizon\footnote{Description of notation: $v$ and $u$ denote in-going and out-going modes, respectively. Outer horizon and cosmological horizon are denoted by $+$ and $c$, respectively. For instance, $ \lmodem(r_*)$ denotes in-going mode at outer horizon.},
\begin{equation}\label{radialmodes}
\begin{split}
\lmodem(r_*) &= {1\over\sqrt{2\omega}}\begin{cases}
e^{-i\omega r_*} & r_*\rightarrow- \infty \equiv r\rightarrow r_+\\
\boga e^{-i\omega r_*} + \bogb e^{i\omega r_*}& r_*\rightarrow \infty \equiv r\rightarrow r_c
\end{cases}\\
\rmodem(r_*) &= {1\over\sqrt{2\omega}}\begin{cases}
e^{i\omega r_*} & r_*\rightarrow- \infty \equiv r\rightarrow r_+\\
\bogas e^{i\omega r_*} + \bogbs e^{-i\omega r_*}& r_*\rightarrow \infty \equiv r\rightarrow r_c
\end{cases}\\
\lmodep(r_*) &= {1\over\sqrt{2\omega}}\begin{cases}
e^{-i\omega r_*} & r_*\rightarrow \infty\equiv r\rightarrow r_c\\
\bogas e^{-i\omega r_*} - \bogb e^{i\omega r_*}& r_*\rightarrow -\infty \equiv r\rightarrow r_+
\end{cases}\\
\rmodep(r_*) &= {1\over\sqrt{2\omega}}\begin{cases}
e^{i\omega r_*} & r_*\rightarrow \infty \equiv r\rightarrow r_c\\
\boga e^{i\omega r_*} - \bogbs e^{-i\omega r_*}& r_*\rightarrow -\infty \equiv r\rightarrow r_+
\end{cases}
\end{split}
\end{equation}
The Bogoliubov coefficients, $\boga$ and $\bogb$ can be determined by solving the radial wave equation \eqref{radialwave}. In third and fourth equation above, we have used $|\boga|^2-|\bogb|^2 = 1$, which can be checked by computing the Wronskian. Using these radial modes, we define the following expansion of the scalar field,
\begin{equation}\label{modeexpsigmap}
\phi = \sum_{\vec{\ell}\,} \int d\omega  \left[a_{\omega,\vec{\ell}\,}\, \rpinmode_{\omega,\vec{\ell}\,}(t,r_*,\Omega) + b_{\omega,\vec{\ell}\,}\,\rcoutmode_{\omega,\vec{\ell}\,}(t,r_*,\Omega)\right] + \text{h.c.}\,.
\end{equation}
where,
\begin{equation}
\begin{split}
\rpinmode_{\omega,\vec{\ell}\,}(t,r_*,\Omega) = {\normalization_{\omega,\ell}\over \sqrt{2\pi} \, r^{d-1\over2} }  \lmodem(r_*) e^{-i\omega t}Y_{\vec{\ell}\,}(\Omega),\\
\rcoutmode_{\omega,\vec{\ell}\,}(t,r_*,\Omega) = {\normalization_{\omega,\ell}\over \sqrt{2\pi} \, r^{d-1\over2} }  \rmodep(r_*) e^{-i\omega t}Y_{\vec{\ell}\,}(\Omega).
\end{split}
\end{equation}
The mode $\rpinmode$ represents an in-going wave at the future inner horizon and $\rcoutmode$ represents an out going wave at the future cosmological horizon. We can also define another mode expansion, using in-going and out-going modes at past cosmological horizon and past outer horizon, respectively.
\begin{equation}\label{modeexpsigmam}
\phi = \sum_{\vec{\ell}\,} \int {d\omega \over\sqrt{2\pi}}{\normalization_{\omega,\ell} \over r^{d-1\over2} } \left[c_{\omega,\vec{\ell}\,}\, \lmodep(r_*) + d_{\omega,\vec{\ell}\,}\,\rmodem(r_*)\right]e^{- i \omega t} Y_{\vec{\ell}\,}(\Omega) + \text{h.c.}\,.
\end{equation}
The Bogoliubov transformation relating the two expansions can be found using \eqref{radialmodes}.
\begin{equation}\label{bogexterior}
\begin{split}
c_{\omega,\vec{\ell}\,} &=  {a_{\omega,\vec{\ell}\,}\over \bogas}- {b_{\omega,\vec{\ell}\,} \bogbs \over \bogas},\\
d_{\omega,\vec{\ell}\,} &= {b_{\omega,\vec{\ell}\,}\over \bogas} + {a_{\omega,\vec{\ell}\,} \bogb \over \bogas}.
\end{split}
\end{equation}
The normalization can be fixed by computing the Klein-Gordon inner product, \eqref{innernull}. For non-zero frequency we can check that,
\begin{equation}
\begin{split}
\left( \rpinmode _{\omega,\vec{\ell}\,} , \rpinmode_{\omega,\vec{\ell}\,'}\right) &= {|A_{\omega,\ell}|^2 |\normalization_{\omega,\ell}|^2}\delta(\omega-\omega')\delta_{\vec{\ell}\,,\vec{\ell}\,'},\\
\left( \rcoutmode _{\omega,\vec{\ell}\,} , \rcoutmode_{\omega,\vec{\ell}\,'}\right) &= {|A_{\omega,\ell}|^2 |\normalization_{\omega,\ell}|^2}\delta(\omega-\omega')\delta_{\vec{\ell}\,,\vec{\ell}\,'},\\
\left( \rpinmode _{\omega,\vec{\ell}\,} , \rcoutmode_{\omega,\vec{\ell}\,'}\right) &= 0.
\end{split}
\end{equation}
For canonical normalization, we choose,
\begin{equation}
\normalization_{\omega,\ell} = {1\over \boga}.
\end{equation}
With this choice, the creation and annihilation operators satisfy the following commutation relations,
\begin{equation}\label{commutatorglobal}
[a_{\omega,\vec{\ell}\,},a^{\dagger}_{\omega',\vec{\ell}\,'}] =[b_{\omega,\vec{\ell}\,},b^{\dagger}_{\omega',\vec{\ell}\,'}] =[c_{\omega,\vec{\ell}\,},c^{\dagger}_{\omega',\vec{\ell}\,'}] =[d_{\omega,\vec{\ell}\,},d^{\dagger}_{\omega',\vec{\ell}\,'}] = \delta(\omega-\omega') \delta_{\vec{\ell}\,,\vec{\ell}\,'}.
\end{equation}
We now study the behaviour of the scalar field near various horizons.
\paragraph{Mode expansion just outside outer horizon\\}
To determine the expansion of scalar field just outside the outer horizon, we start with \eqref{modeexpsigmap} and use \eqref{radialmodes}.
\begin{equation}\label{modeoutsideouter}
\phi \xrightarrow[ r\rightarrow r_+ ]{} \sum_{\vec{\ell}\,} \int {d\omega \over\sqrt{2\pi}}{\left[\left(a_{\omega,\vec{\ell}\,} -\bogbs\, b_{\omega,\vec{\ell}\,}\right) e^{- i \omega r_*} + \boga\, b_{\omega,\vec{\ell}\,} e^{i\omega r_*}\right]\over \boga\sqrt{2\omega}r^{d-1\over2} } e^{- i \omega t} Y_{\vec{\ell}\,}(\Omega) +\text{h.c.}\, .
\end{equation}
We can use the Kruskal coordinates that are well defined across the outer horizon, \eqref{kruksalp}, to recast the expansion as
\begin{equation}\label{modeoutsideouterkruskal}
\phi \xrightarrow[ r\rightarrow r_+ ]{} \sum_{\vec{\ell}\,} \int {d\omega \over\sqrt{2\pi}}\left[ {\bogas\over \boga} c_{\omega,\vec{\ell}\,}  \left({\kappa_+\over\zeta_+}V_+ \right)^{-i\omega\over\kappa_+} +  b_{\omega,\vec{\ell}\,} \left({-\kappa_+\over\zeta_+}U_+ \right)^{i\omega\over\kappa_+}\right] {Y_{\vec{\ell}\,}(\Omega) \over \sqrt{2\omega}r^{d-1\over2}  } +\text{h.c.}\, ,
\end{equation}
where we have used \eqref{bogexterior} to simplify the expression.
\paragraph{Mode expansion just inside cosmological horizon\\}
Just inside the future cosmological horizon, scalar field mode expansion takes the following form.
\begin{equation}\label{modefuturecosmological}
\phi \xrightarrow[ r\rightarrow r_c ]{} \sum_{\vec{\ell}\,} \int {d\omega \over\sqrt{2\pi}}{\left[\left(b_{\omega,\vec{\ell}\,} +\bogb\, a_{\omega,\vec{\ell}\,}\right) e^{ i \omega r_*} + \boga\, a_{\omega,\vec{\ell}\,} e^{-i\omega r_*}\right]\over \boga\sqrt{2\omega}r^{d-1\over2} } e^{- i \omega t} Y_{\vec{\ell}\,}(\Omega) +\text{h.c.}\, .
\end{equation}
Just as before, using Kruskal coordinate regular at cosmological horizon \eqref{kruksalc}, we can recast this expansion as
\begin{equation}\label{modeinsidecosmologicalkruskal}
\phi \xrightarrow[ r\rightarrow r_c ]{} \sum_{\vec{\ell}\,} \int {d\omega \over\sqrt{2\pi}}\left[ {\bogas\over \boga} d_{\omega,\vec{\ell}\,}  \left({\kappa_c\over\zeta_c}U_c \right)^{-i\omega\over\kappa_c} +  a_{\omega,\vec{\ell}\,} \left({-\kappa_c\over\zeta_c}V_c \right)^{i\omega\over\kappa_c}\right] {Y_{\vec{\ell}\,}(\Omega) \over \sqrt{2\omega}r^{d-1\over2}  } +\text{h.c.}\, ,
\end{equation}

\subsection{Local modes near horizon in the exterior}
Following \cite{Papadodimas:2019msp}, we define the local modes near the horizons by integrating the scalar field in appropriate Kruskal coordinates introduced section \ref{s_geometry}. Since solutions of the Klein-Gordon equation with fixed energy oscillates infinitely near horizons, local modes of a particular frequency can be extracted by integrating only in a neighbourhood of a horizon. Due to spherical symmetry, we can extract near horizon modes for each angular momentum as well. 

We define a real ``tuning'' function, $\tune(x)$, which has support only very close to zero, $x\in[x_l,x_h]$ and $0 < x_l \ll x_0 \ll x_h \ll R$, where $R$ denotes the characteristic curvature scale outside the inner horizon.
\begin{equation}\label{sharpdef}
\tune(x)   = \int_{-\infty}^{\infty} \sharp(\nu) \left({x \over x_0} \right)^{i \nu} d \nu; \qquad \sharp(\nu)   = {1 \over 2 \pi} \int_{0}^{\infty} {d x \over x} \tune(x) \left({x \over x_0} \right)^{-i \nu}. 
\end{equation}
Also, we normalize the tuning function by demanding
\begin{equation}\label{normtuning}
\int   |\sharp(\nu)|^2 {d \nu} = 1.
\end{equation}
We choose $\sharp(\nu)$ to be  sharply peaked around $\nu = 0$, which corresponds to $\tune(x)$ being almost constant in the domain $[x_l,x_h]$. Since the tuning function is real, $\sharp(-\nu)=\sharp(\nu)$.

\begin{figure}
	\centering
	\includegraphics[width=6cm]{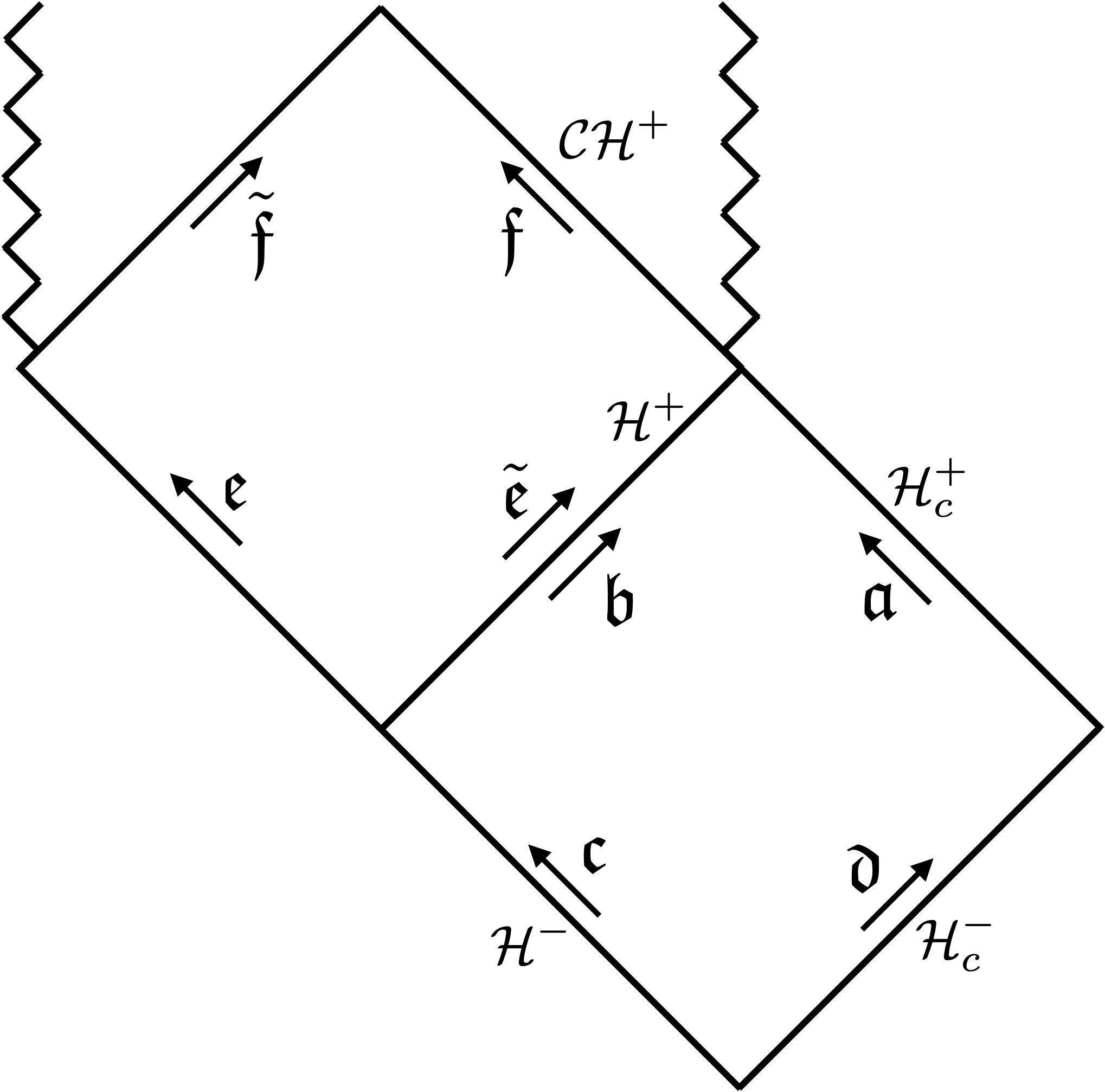}
	\caption{A graphic to illustrate the various near horizon modes defined throughout the paper.\label{fig_localmodes}}
\end{figure}

Now, we define various near horizon modes in the exterior region as follows.
\begin{equation}\label{localmodesextdef}
\begin{split}
\anha_{\omega,\vec{\ell}\,} &=  {r_c^{d-1\over2} \sqrt{\kappa_c} \over \sqrt{\pi \omega}} \int \partial_{V_c} \phi (U_c = \epsilon, V_c ,\Omega) \left({-V_c\over x_0} \right)^{-i{\omega\over\kappa_c}} \tune(-V_c)Y^*_{\vec{\ell}\,}(\Omega)dV_c d^{d-1}\Omega,\\
\anhb_{\omega,\vec{\ell}\,} &=  {r_+^{d-1\over2} \sqrt{\kappa_+}\over \sqrt{\pi \omega}} \int \partial_{U_+} \phi (U_+, V_+= \epsilon ,\Omega) \left({-U_+\over x_0} \right)^{-i{\omega\over\kappa_+}} \tune(-U_+)Y^*_{\vec{\ell}\,}(\Omega)dU_+ d^{d-1}\Omega,\\
\anhc_{\omega,\vec{\ell}\,} &=  {r_+^{d-1\over2} \sqrt{\kappa_+} \over \sqrt{\pi \omega}} \int \partial_{V_+} \phi (U_+ = -\epsilon, V_+ ,\Omega) \left({V_+\over x_0} \right)^{i{\omega\over\kappa_+}} \tune(V_+)Y^*_{\vec{\ell}\,}(\Omega)dV_+ d^{d-1}\Omega,\\
\anhd_{\omega,\vec{\ell}\,} &=  {r_c^{d-1\over2} \sqrt{\kappa_c} \over \sqrt{\pi \omega}} \int \partial_{U_c} \phi (U_c , V_c = -\epsilon ,\Omega) \left({U_c\over x_0} \right)^{i{\omega\over\kappa_c}} \tune(U_c)Y^*_{\vec{\ell}\,}(\Omega)dU_c d^{d-1}\Omega.
\end{split}
\end{equation}
The phases in the integral are chosen to extract positive energy modes with frequency $\omega$. Integral over the sphere with spherical harmonics ensures that we extract modes with angular momentum $\vec{\ell}$. The normalization ensures that local operators satisfy following commutation relations.
\begin{equation}\label{commutationlocal}
[\anha_{\omega,\vec{\ell}\,},\anha_{\omega,\vec{\ell}\,}^{\dagger}] = [\anhb_{\omega,\vec{\ell}\,},\anhb_{\omega,\vec{\ell}\,}^{\dagger}] =[\anhc_{\omega,\vec{\ell}\,},\anhc_{\omega,\vec{\ell}\,}^{\dagger}] =[\anhd_{\omega,\vec{\ell}\,},\anhd_{\omega,\vec{\ell}\,}^{\dagger}] =1.
\end{equation}
As the tuning function has support only for small and positive arguments, we can use the mode expansion of scalar field near various horizons, \eqref{modeoutsideouterkruskal} and \eqref{modeinsidecosmologicalkruskal}, to simplify the integrals. This leads to following relation between local and global modes. 
\begin{equation}\label{localglobalmodesrelation}
\begin{split}
\anha_{\omega,\vec{\ell}\,} &=  i \int \sharp( {\omega -\omega'\over\kappa_c}) ({\kappa_c x_0\over\zeta_c})^{i\omega'\over\kappa_c}\sqrt{\kappa_c\omega'\over\omega} a_{\omega',\vec{\ell}\,} {d\omega'\over\kappa_c},\\
\anhb_{\omega,\vec{\ell}\,} &= i \int \sharp( {\omega -\omega'\over\kappa_+}) ({\kappa_+ x_0\over\zeta_+})^{i\omega'\over\kappa_+}\sqrt{\kappa_+\omega'\over\omega} b_{\omega',\vec{\ell}\,} {d\omega'\over\kappa_+},\\
\anhc_{\omega,\vec{\ell}\,} &=  - i \int \sharp( {\omega -\omega'\over\kappa_+}) ({\kappa_+ x_0\over\zeta_+})^{-i\omega'\over\kappa_+}\sqrt{\kappa_+\omega'\over\omega} {\bogal^*_{\omega',\vec{\ell}}\over\bogal_{\omega,\vec{\ell}}} c_{\omega',\vec{\ell}\,} {d\omega'\over\kappa_+},\\
\anhd_{\omega,\vec{\ell}\,} &=  - i \int \sharp( {\omega -\omega'\over\kappa_c}) ({\kappa_c x_0\over\zeta_c})^{-i\omega'\over\kappa_c}\sqrt{\kappa_c\omega'\over\omega} {\bogal^*_{\omega',\vec{\ell}}\over\bogal_{\omega,\vec{\ell}}} d_{\omega',\vec{\ell}\,} {d\omega'\over\kappa_c}.
\end{split}
\end{equation}
Using the commutators of the global modes, \eqref{commutatorglobal} and properties of the tuning function \eqref{normtuning}, it is easy to check that \eqref{commutationlocal} is satisfied\footnote{Recall that the function $\sharp$ is sharply peaked around zero.}. As the integrals in \eqref{localmodesextdef} were restricted to a finite region, we could only extract global modes smeared over a range of frequencies. However, since mode functions of the scalar field oscillates infinitely near horizons, we were able to restrict the smearing to a very small range of frequencies. The local modes $\anha$, $\anhb$, $\anhc$ and $\anhd$ are \emph{slightly smeared} version of global modes that are in-going at cosmological horizon, out-going at outer horizon, in-going at outer horizon and out-going at cosmological horizon, respectively. In Figure \ref{fig_localmodes}, we illustrate all near horizon modes for clarity\footnote{The figure also includes near horizon modes in the interior, to be defined later.}. 

\subsection{Smooth exterior}
Having defined the near horizon modes in the exterior, we now demonstrate the existence of a quantum state, $|\Psi\rangle$, that is regular everywhere in this region. Suppose, the state $|\Psi\rangle$ is smooth as we approach the future outer horizon. Then, the two point function of scalar field with insertions close to the future outer horizon should satisfy the following limit.
\begin{equation}\label{twoptfn}
\lim\limits_{x_1 \rightarrow x_2} \langle \Psi | \phi(x_1)\phi(x_2)|\Psi\rangle =  {\Gamma(d-1) \over 2^d \pi^{d \over2} \Gamma({d \over 2})}{1 \over s^{d-1 \over 2}} + {\cal R}.
\end{equation}
In the above equation, $x_i$ is the spacetime coordinate of $i^{th}$ insertion and $s$ is the geodesic distance between $x_1$ and $x_2$. Since the insertions are close to the outer horizon, the geodesic distance can be computed using the metric \eqref{metricp}. ${\cal R}$ denotes terms which are sub-leading as $x_1$ approaches $x_2$. The above equation is simply the flat-space limit of the two-point function of the scalar field. In computation of the renormalized stress tensor, we subtract the contribution of this term to get a finite answer. If \eqref{twoptfn} is not satisfied, then the stress tensor near the future outer horizon would diverge.

In \cite{Papadodimas:2019msp}, it was shown that \eqref{twoptfn} is enough to fix the two point functions of near horizon modes. Regularity of the state $|\Psi\rangle$ as we approach the future leg of outer horizon from the outside fixes the two-point function of near horizon mode $\anhb$.
\begin{equation}\label{bbdagger}
\langle\Psi|\anhb_{\omega,\vec{\ell}\,}\anhb_{\omega,\vec{\ell}\,}^{\dagger}|\Psi\rangle = {1\over 1- e^{- {2\pi \omega \over \kappa_+}}}.
\end{equation}
The modes $\anhb$ are occupied thermally, with the characteristic temperature of the outer horizon, $\kappa_{+}\over 2\pi$. This is the familiar Hawking radiation emanating from the outer horizon. The thermal occupancy of near horizon modes constrains the two point function of global modes via the relation \eqref{localglobalmodesrelation}. To see this, we start with the following ansatz for the two point function of global modes,
\begin{equation}
\langle\Psi|b_{\omega,\vec{\ell}\,}b^{\dagger}_{\omega',\vec{\ell}\,}|\Psi\rangle = \mathfrak{B}(\omega)\delta(\omega-\omega') + \tilde{\mathfrak{B}}(\omega,\omega').
\end{equation}
In the above expression, the function $\tilde{\mathfrak{B}}$ is assumed to be regular at $\omega=\omega'$. Since the near horizon modes are obtained by smearing the global modes with functions that have support in a very small range of frequency, $\tilde{\mathfrak{B}}$ does not contribute to the two-point function of near horizon modes, \eqref{bbdagger}. However, a simple calculation shows that the form of $\mathfrak{B}$ is \emph{completely fixed}.
\begin{equation}
\mathfrak{B}(\omega) = {1\over 1- e^{- {2\pi \omega \over \kappa_+}}} .
\end{equation}
In anti-de-Sitter space, due to normalizable boundary conditions, the in-going and out-going modes are identified. Hence, smoothness of outer horizon is sufficient to fix all two-point functions involving near horizon modes. However, in de-Sitter space, apart from $\langle\Psi|\anhb\anhb^{\dagger}|\Psi\rangle$, we also need to fix $\langle\Psi|\anha\anha^{\dagger}|\Psi\rangle$ and $\langle\Psi|\anhb\anha^{\dagger}|\Psi\rangle$. 

For that, we assume smoothness of quantum state at the future and past cosmological horizon as well. This physical assumption fixes the two point function of near horizon modes $\anha$ and $\anhd$,
\begin{equation}\label{aadagger}
\langle\Psi|\anha_{\omega,\vec{\ell}\,}\anha_{\omega,\vec{\ell}\,}^{\dagger}|\Psi\rangle = {1\over 1- e^{- {2\pi \omega \over \kappa_c}}},
\end{equation}
and 
\begin{equation}\label{dddagger}
\langle\Psi|\anhd_{\omega,\vec{\ell}\,}\anhd_{\omega,\vec{\ell}\,}^{\dagger}|\Psi\rangle = {1\over 1- e^{- {2\pi \omega \over \kappa_c}}}.
\end{equation}
Once again, this constrains the two-point function of global modes. We collect all two-point functions of global modes fixed by assuming smoothness of future outer horizon, and future and past cosmological horizon.
\begin{equation}\label{2pointglobal}
\begin{split}
\langle\Psi|b_{\omega,\vec{\ell}\,}b_{\omega',\vec{\ell}\,}^{\dagger}|\Psi\rangle = {1\over 1- e^{- {2\pi \omega \over \kappa_+}}} \delta(\omega-\omega') + \cdots,\\
\langle\Psi|a_{\omega,\vec{\ell}\,}a_{\omega',\vec{\ell}\,}^{\dagger}|\Psi\rangle = {1\over 1- e^{- {2\pi \omega \over \kappa_c}}} \delta(\omega-\omega')  + \cdots,\\
\langle\Psi|d_{\omega,\vec{\ell}\,}d_{\omega',\vec{\ell}\,}^{\dagger}|\Psi\rangle = {1\over 1- e^{- {2\pi \omega \over \kappa_c}}} \delta(\omega-\omega')  + \cdots,
\end{split}
\end{equation}
where ellipsis denote terms regular at $\omega=\omega'$. Together with the Bogoliubov transformation in \eqref{bogexterior}, the above equation fixes the following two-point functions as well.
\begin{equation}\label{abdaggerglobal}
\begin{split}
\langle\Psi|a_{\omega,\vec{\ell}\,}b_{\omega',\vec{\ell}\,}^{\dagger}|\Psi\rangle = \langle\Psi|b_{\omega,\vec{\ell}\,}a_{\omega',\vec{\ell}\,}^{\dagger}|\Psi\rangle 
&= {\delta(\omega-\omega')\over \bogb +\bogbs} \left[  {1\over 1- e^{- {2\pi \omega \over \kappa_c}}} - {1\over 1- e^{- {2\pi \omega \over \kappa_+}}}  \right]+ \cdots \,.
\end{split}
\end{equation}
We can use this to determine,
\begin{equation}\label{abdagger}
\langle\Psi| \anha_{\omega,\vec{\ell}\,}\anhb_{\omega,\vec{\ell}\,}^{\dagger}|\Psi\rangle = {\left({\kappa_c\zeta_+\over\kappa_+\zeta_c}\right)^{i\omega} \over \bogb +\bogbs } \left[  {1\over 1- e^{- {2\pi \omega \over \kappa_c}}} - {1\over 1- e^{- {2\pi \omega \over \kappa_+}}}  \right]
\end{equation}

\paragraph{Smoothness of past outer horizon \\}
Having fixed all two-point function of local modes in the exterior, we now compute the occupation of the local modes near the past outer horizon. To do that, we first compute the two-point function of global mode $c$ using the Bogoliubov transformation \eqref{bogexterior}, and the two-point functions \eqref{2pointglobal} and \eqref{abdaggerglobal}. We find that,
\begin{equation}\label{eedaggerglobal}
\langle\Psi|c_{\omega,\vec{\ell}\,}c_{\omega',\vec{\ell}\,}^{\dagger}|\Psi\rangle = {1\over 1- e^{- {2\pi \omega \over \kappa_+}}} \delta(\omega-\omega')+ \cdots.
\end{equation}
Using the definition of $\anhc$, \eqref{localglobalmodesrelation}, this immediately implies that the local modes near past outer horizon are thermally populated with the characteristic temperature of the outer horizon. 
\begin{equation}\label{ccdagger}
\langle\Psi|\anhc_{\omega,\vec{\ell}\,}\anhc_{\omega,\vec{\ell}\,}^{\dagger}|\Psi\rangle = {1\over 1- e^{- {2\pi \omega \over \kappa_+}}},
\end{equation}
This is precisely what we expect in a state which is regular at the past outer horizon. The assumption of regularity of the quantum state at future outer horizon, and future and past cosmological horizon automatically ensures regularity at the past outer horizon. This strongly suggests the possibility of existence of quantum states with finite stress tensor everywhere in the exterior. This may seem surprising as the cosmological horizon and outer horizon radiate at different temperatures.



\section{Quantum instability of the inner horizon \label{s_interior}}
\subsection{Modes expansion in the interior}
In the interior region, i.e., between the inner and outer horizon, we define the following solutions to the radial wave equation, by fixing the behavior near the horizons\footnote{We have, once again, used $v$ and $u$ to denote the in-going and out-going mode, and $+,-$ to denote the outer and inner horizon. The \~{} indicates that these solutions are valid in the interior.},
\begin{equation}\label{radialmodeinsidebh}
\begin{split}
\lmodeinsidep(r_*) &= {1\over\sqrt{2\omega}}\begin{cases}
e^{-i\omega r_*} & r_*\rightarrow- \infty \equiv r \rightarrow r_+\\
\bogainside e^{-i\omega r_*} + \bogbinside e^{i\omega r_*}& r_*\rightarrow \infty\equiv r \rightarrow r_-
\end{cases}\\
\rmodeinsidep(r_*) &= {1\over\sqrt{2\omega}}\begin{cases}
e^{i\omega r_*} & r_*\rightarrow- \infty\equiv r \rightarrow r_+\\
\bogasinside e^{i\omega r_*} + \bogbsinside e^{-i\omega r_*}& r_*\rightarrow \infty\equiv r \rightarrow r_-
\end{cases}\\
\lmodeinsidem(r_*) &= {1\over\sqrt{2\omega}}\begin{cases}
e^{-i\omega r_*} & r_*\rightarrow \infty\equiv r \rightarrow r_- \\
\bogasinside e^{-i\omega r_*} - \bogbinside e^{i\omega r_*}& r_*\rightarrow- \infty\equiv r \rightarrow r_+
\end{cases}\\
\rmodeinsidem(r_*) &= {1\over\sqrt{2\omega}}\begin{cases}
e^{i\omega r_*} & r_*\rightarrow \infty\equiv r \rightarrow r_-\\
\bogainside e^{i\omega r_*} - \bogbsinside e^{-i\omega r_*}&  r_*\rightarrow- \infty\equiv r \rightarrow r_+
\end{cases}
\end{split}
\end{equation}
where, the Bogoliubov coefficients can be obtained by solving the radial wave equation inside the black holes. Once again, the conservation of Wronskian implies $|\bogainside|^2-|\bogbinside|^2 = 1$.

For each frequency and angular momentum, there are two independent modes in the interior. The in-going mode is obtained by continuing the in-going mode from the exterior. However, due to the causal structure of the spacetime, we need to define a new outgoing mode.

We define first set of modes in the interior via the following field expansion.
\begin{equation}\label{modeexpinteriorp}
\phi = \sum_{\vec{\ell}\,} \int {d\omega \over\sqrt{2\pi}}{1\over r^{d-1\over2}} \left[e_{\omega,\vec{\ell}\,}\, \lmodeinsidep(r_*) + \tilde{e}^{\dagger}_{\omega,\vec{\ell}\,}\,\rmodeinsidep(r_*)\right]e^{- i \omega t} Y_{\vec{\ell}\,}(\Omega) + \text{h.c.}\,.
\end{equation}
Using \eqref{radialmodeinsidebh}, it is clear that the above expansion takes a simple form near the outer horizon. Since $r_*$ is the time coordinate inside the black hole, $\rmodeinsidep$ corresponds to a negative energy mode near the outer horizon. This justifies the association of a \emph{creation} operator with this mode.

We also define another set of modes by the expansion,
\begin{equation}\label{modeexpinteriorm}
\phi = \sum_{\vec{\ell}\,} \int {d\omega \over\sqrt{2\pi}}{1\over r^{d-1\over2}} \left[f_{\omega,\vec{\ell}\,}\, \lmodeinsidem(r_*) + \tilde{f}^{\dagger}_{\omega,\vec{\ell}\,}\,\rmodeinsidem(r_*)\right]e^{- i \omega t} Y_{\vec{\ell}\,}(\Omega) + \text{h.c.}\,.
\end{equation}
Using \eqref{radialmodeinsidebh}, we obtain the Bogoliubov transformation relating the two sets of modes,
\begin{equation}\label{boginterior}
\begin{split}
f_{\omega,\vec{\ell}\,} &= \bogainside \, e_{\omega,\vec{\ell}\,} + \bogbsinside \, \tilde{e}^{\dagger}_{\omega,\vec{\ell}\,},\\
\tilde{f}_{\omega,\vec{\ell}\,} &= \bogainside \,\tilde{e}_{\omega,\vec{\ell}\,} +\bogbsinside \, e^{\dagger}_{\omega,\vec{\ell}\,}.
\end{split}
\end{equation}
Note that the Bogoliubov transformation now mixes positive and negative energy modes. This is expected as time translation inside the black hole, i.e., translations in coordinate $r_*$, is not a symmetry. The normalization has been chosen such that 
\begin{equation}\label{commutatorglobalinterior}
[e_{\omega,\vec{\ell}\,},e^{\dagger}_{\omega',\vec{\ell}\,'}] =[\tilde{e}_{\omega,\vec{\ell}\,},\tilde{e}^{\dagger}_{\omega',\vec{\ell}\,'}] =[f_{\omega,\vec{\ell}\,},f^{\dagger}_{\omega',\vec{\ell}\,'}] =[\tilde{f}_{\omega,\vec{\ell}\,},\tilde{f}^{\dagger}_{\omega',\vec{\ell}\,'}] = \delta(\omega-\omega') \delta_{\vec{\ell}\,,\vec{\ell}\,'}.
\end{equation}
To see how in-going modes in the interior are related to that in the exterior, we expand the scalar field near the outer horizon.
\paragraph{Mode expansion just inside the outer horizon\\}
Using \eqref{modeexpinteriorp}, the scalar field can be expanded just inside the outer horizon as follows.
\begin{equation}\label{modeexpinsideouter}
\phi \xrightarrow[ r\rightarrow r_+ ]{} \sum_{\vec{\ell}\,} \int {d\omega \over\sqrt{2\pi}r^{d-1\over2}}\left[e_{\omega,\vec{\ell}\,}e^{- i \omega r_*}+  \tilde{e}^{\dagger}_{\omega,\vec{\ell}\,} e^{i\omega r_*} \right] {e^{- i \omega t}\over \sqrt{2\omega}} Y_{\vec{\ell}\,}(\Omega) +\text{h.c.}\, ,
\end{equation}
We can use \eqref{kruksalp} to rewrite the above expansion as,
\begin{equation}\label{modeexpinsideouterkruskal}
\phi \xrightarrow[ r\rightarrow r_+ ]{} \sum_{\vec{\ell}\,} \int {d\omega \over\sqrt{2\pi}r^{d-1\over2}}\left[e_{\omega,\vec{\ell}\,}\left({\kappa_+\over\zeta_+}V_+ \right)^{-i\omega\over\kappa_+} +  \tilde{e}^{\dagger}_{\omega,\vec{\ell}\,} \left({\kappa_+\over\zeta_+}U_+ \right)^{i\omega\over\kappa_+} \right] {Y_{\vec{\ell}\,}(\Omega)\over \sqrt{2\omega}}  +\text{h.c.}\, ,
\end{equation}
Comparing with \eqref{modeoutsideouterkruskal} and using continuity of the field, we can relate in-going modes across the outer horizon.
\begin{equation}\label{globalingoingmodeacrosshorizon}
e_{\omega,\vec{\ell}\,} = {\bogas\over \boga} c_{\omega,\vec{\ell}\,}.
\end{equation}
\paragraph{Mode expansion just outside the inner horizon\\}
Using \eqref{modeexpinteriorm} the scalar field can be expanded just outside the inner horizon as follows.
\begin{equation}\label{modeexpoutsideinner}
\phi \xrightarrow[ r\rightarrow r_- ]{} \sum_{\vec{\ell}\,} \int {d\omega \over\sqrt{2\pi}r^{d-1\over2}}\left[f_{\omega,\vec{\ell}\,}e^{- i \omega r_*}+  \tilde{f}^{\dagger}_{\omega,\vec{\ell}\,} e^{i\omega r_*} \right] {e^{- i \omega t}\over \sqrt{2\omega}} Y_{\vec{\ell}\,}(\Omega) +\text{h.c.}\, ,
\end{equation}
We can use \eqref{kruksalm} to rewrite the above expansion as,
\begin{equation}\label{modeexpoutsideinnerkruskal}
\phi \xrightarrow[ r\rightarrow r_- ]{} \sum_{\vec{\ell}\,} \int {d\omega \over\sqrt{2\pi}r^{d-1\over2}}\left[f_{\omega,\vec{\ell}\,}\left(-{\kappa_-\over\zeta_-}V_- \right)^{i\omega\over\kappa_-} +  \tilde{f}^{\dagger}_{\omega,\vec{\ell}\,} \left(-{\kappa_-\over\zeta_-}U_- \right)^{-i\omega\over\kappa_-} \right] {Y_{\vec{\ell}\,}(\Omega)\over \sqrt{2\omega}}  +\text{h.c.}\, ,
\end{equation}

\subsection{Local modes near horizon in the interior}
Following the procedure in previous section, we define the near horizon modes in the interior.
\begin{equation}\label{localmodesintdef}
\begin{split}
\anhe_{\omega,\vec{\ell}\,} &=  {r_+^{d-1\over2} \sqrt{\kappa_+} \over \sqrt{\pi \omega}} \int \partial_{V_+} \phi (U_+ = \epsilon, V_+ ,\Omega) \left({V_+\over x_0} \right)^{i{\omega\over\kappa_+}} \tune(V_+)Y^*_{\vec{\ell}\,}(\Omega)dV_+ d^{d-1}\Omega,\\
\anhet_{\omega,\vec{\ell}\,} &=  {r_+^{d-1\over2} \sqrt{\kappa_+}\over \sqrt{\pi \omega}} \int \partial_{U_+} \phi (U_+, V_+= \epsilon ,\Omega) \left({U_+\over x_0} \right)^{i{\omega\over\kappa_+}} \tune(U_+)Y_{\vec{\ell}\,}(\Omega)dU_+ d^{d-1}\Omega,\\
\anhf_{\omega,\vec{\ell}\,} &=  {r_-^{d-1\over2} \sqrt{\kappa_-} \over \sqrt{\pi \omega}} \int \partial_{V_-} \phi (U_- = -\epsilon, V_- ,\Omega) \left(-{V_-\over x_0} \right)^{-i{\omega\over\kappa_-}} \tune(-V_-)Y^*_{\vec{\ell}\,}(\Omega)dV_- d^{d-1}\Omega,\\
\anhft_{\omega,\vec{\ell}\,} &=  {r_-^{d-1\over2} \sqrt{\kappa_-}\over \sqrt{\pi \omega}} \int \partial_{U_-} \phi (U_-, V_-= -\epsilon ,\Omega) \left(-{U_-\over x_0} \right)^{-i{\omega\over\kappa_-}} \tune(-U_-)Y_{\vec{\ell}\,}(\Omega)dU_- d^{d-1}\Omega,.
\end{split}
\end{equation}
Using the mode expansion \eqref{modeexpinsideouterkruskal} and \eqref{modeexpoutsideinnerkruskal}, we can express the interior near horizon modes as smeared global modes.
\begin{equation}\label{localglobalmodesrelationinterior}
\begin{split}
\anhe_{\omega,\vec{\ell}\,} &=  - i \int \sharp( {\omega -\omega'\over\kappa_+}) ({\kappa_+ x_0\over\zeta_+})^{-i\omega'\over\kappa_+}\sqrt{\kappa_+\omega'\over\omega}  e_{\omega',\vec{\ell}\,} {d\omega'\over\kappa_+},\\
\anhet_{\omega,\vec{\ell}\,} &=  -i \int \sharp( {\omega -\omega'\over\kappa_+}) ({\kappa_+ x_0\over\zeta_+})^{-i\omega'\over\kappa_+}\sqrt{\kappa_+\omega'\over\omega} \tilde{e}_{\omega',\vec{\ell}\,} {d\omega'\over\kappa_c},\\
\anhf_{\omega,\vec{\ell}\,} &=  i \int \sharp( {\omega -\omega'\over\kappa_-}) ({\kappa_- x_0\over\zeta_-})^{i\omega'\over\kappa_-}\sqrt{\kappa_-\omega'\over\omega}  f_{\omega',\vec{\ell}\,} {d\omega'\over\kappa_-},\\
\anhft_{\omega,\vec{\ell}\,} &=  i \int \sharp( {\omega -\omega'\over\kappa_-}) ({\kappa_- x_0\over\zeta_-})^{i\omega'\over\kappa_-}\sqrt{\kappa_-\omega'\over\omega} \tilde{f}_{\omega',\vec{\ell}\,} {d\omega'\over\kappa_-}.
\end{split}
\end{equation} 
We notice that, as expected, the in-going mode just inside the outer horizon is identical to the in-going mode just outside, see \eqref{globalingoingmodeacrosshorizon} and \eqref{localglobalmodesrelation}. 
\begin{equation}
\anhe_{\omega,\vec{\ell}\,} =\anhc_{\omega,\vec{\ell}\,} .
\end{equation}
As already discussed, this is a consequence of continuity of the field and regularity of the in-going mode function across the future outer horizon. Once again, it is easy to check that the near horizon modes are normalized such that,
\begin{equation}\label{commutationlocalinterior}
[\anhe_{\omega,\vec{\ell}\,},\anhe_{\omega,\vec{\ell}\,}^{\dagger}] = [\anhet_{\omega,\vec{\ell}\,},\anhet_{\omega,\vec{\ell}\,}^{\dagger}] =[\anhf_{\omega,\vec{\ell}\,},\anhf_{\omega,\vec{\ell}\,}^{\dagger}] =[\anhft_{\omega,\vec{\ell}\,},\anhft_{\omega,\vec{\ell}\,}^{\dagger}] =1.
\end{equation}

\subsection{Quantum instability of the inner horizon}
Having defined all the near horizon modes, we now proceed to explore whether the quantum state $|\Psi\rangle$ is smooth in the interior. First, we impose smoothness of quantum state across outer horizon to fix all two point functions of near horizon modes. Then, we show that the local modes near inner horizon are \emph{not} thermally populated with the characteristic temperature of inner horizon, $\kappa_-\over 2\pi$. As a consequence, the expectation value of the quantum stress tensor in state $|\Psi\rangle$ diverges at the inner horizon. Hence, quantum fluctuations restore the strong cosmic censorship conjecture.

\subsubsection{Smoothness of state just inside the outer horizon}
First, we note that the two point function of scalar field has the correct flat space limit \eqref{twoptfn} as the insertions approach the left leg of future outer horizon, $V_+\rightarrow0^+$. This is the case since the in-going modes near the future horizon are thermally populated with the characteristic temperature of the outer horizon, see \eqref{ccdagger}.
\begin{equation}\label{eedagger}
\langle\Psi|\anhe_{\omega,\vec{\ell}\,}\anhe_{\omega,\vec{\ell}\,}^{\dagger}|\Psi\rangle =\langle\Psi|\anhc_{\omega,\vec{\ell}\,}\anhc_{\omega,\vec{\ell}\,}^{\dagger}|\Psi\rangle= {1\over 1- e^{- {2\pi \omega \over \kappa_+}}},
\end{equation}
If we approach the right leg of future outer horizon from the interior, $U_+\rightarrow0^+$, then the smoothness of quantum state would require,
\begin{equation}\label{etetdagger}
\langle\Psi|\anhet_{\omega,\vec{\ell}\,}\anhet_{\omega,\vec{\ell}\,}^{\dagger}|\Psi\rangle = {1\over 1- e^{- {2\pi \omega \over \kappa_+}}}.
\end{equation}

\subsubsection{Entanglement of local modes across the outer horizon}
Smoothness of quantum state as we approach the outer horizon from one side leads to results \eqref{eedagger} and \eqref{etetdagger}. However, this is not sufficient to fix all two-point function of the near horizon modes. The two point function $\langle\Psi| \anhe_{\omega,\vec{\ell}\,} \anhet_{\omega,\vec{\ell}\,}|\Psi\rangle$ remains undetermined. To fix this, we enforce smoothness of the quantum state $|\psi\rangle$ across the outer horizon. We assume that the two point function of the scalar field reduces to the flat space limit \eqref{twoptfn} when the two insertions approach the outer horizon ($U_+=0$) from the opposite sides. The smoothness across the horizon requires \cite{Papadodimas:2019msp}, 
\begin{equation}\label{bet}
\langle\Psi|\anhb_{\omega,\vec{\ell}\,}\anhet_{\omega,\vec{\ell}\,}|\Psi\rangle = {e^{- {\pi \omega \over \kappa_+}}\over 1- e^{- {2\pi \omega \over \kappa_+}}}.
\end{equation}
Consider the action of operator $\anhet$ on the state $|\Psi\rangle$.
\begin{equation}
\anhet_{\omega,\ell}|\Psi\rangle = c_1 \anhb_{\omega,\ell}^{\dagger}|\Psi\rangle + c_2 |\chi \rangle
\end{equation}
where $|\chi\rangle$ is assumed to be orthogonal to the state $\anhb_{\omega,\ell}^{\dagger}|\Psi\rangle$. We fix $c_1$ by taking the inner product of $\anhet_{\omega,\ell}|\Psi\rangle$ with $\anhb_{\omega,\ell}^{\dagger}|\Psi\rangle$. Since  $|\chi\rangle$ is orthogonal to $\anhb_{\omega,\ell}^{\dagger}|\Psi\rangle$,
\begin{equation}
\begin{split}
\langle\Psi|\anhb_{\omega,\vec{\ell}\,}\anhet_{\omega,\vec{\ell}\,}|\Psi\rangle &= {e^{- {\pi \omega \over \kappa_+}}\over 1- e^{- {2\pi \omega \over \kappa_+}}} = c_1 \langle\Psi|\anhb_{\omega,\vec{\ell}\,}\anhb_{\omega,\vec{\ell}\,}^{\dagger}|\Psi\rangle\\
\implies c_1 & = e^{- {\pi \omega \over \kappa_+}}.\
\end{split}
\end{equation}
We have used \eqref{bbdagger} and \eqref{bet} above. To fix $c_2$ we consider the norm of the state $	\anhet_{\omega,\vec{\ell}\,}|\Psi\rangle $.
\begin{equation}
\begin{split}
||	\anhet_{\omega,\vec{\ell}\,}|\Psi\rangle ||^2 &= \langle\Psi| \anhet_{\omega,\vec{\ell}\,}^{\dagger}\anhet_{\omega,\vec{\ell}\,}|\Psi\rangle  = |c_1|^2 \langle\Psi| \anhb_{\omega,\vec{\ell}\,}\anhb_{\omega,\vec{\ell}\,}^{\dagger}|\Psi\rangle + |c_2|^2 \langle\chi|\chi\rangle,\\
&\implies  {e^{- {2\pi \omega \over \kappa_+}}\over 1- e^{- {2\pi \omega \over \kappa_+}}} = {e^{- {2\pi \omega \over \kappa_+}}\over 1- e^{- {2\pi \omega \over \kappa_+}}} + |c_2|^2,\\
&\implies c_2 = 0.
\end{split}
\end{equation}
Hence, the action of out-going local mode behind the outer horizon $\anhet_{\omega,\vec{\ell}\,}$ on the state $|\Psi\rangle$ is completely determined by the action of local mode outside the outer horizon  $\anhb_{\omega,\vec{\ell}\,}$,
\begin{equation}\label{actionet}
\anhet_{\omega,\vec{\ell}\,}|\Psi\rangle = e^{-{\pi \omega\over \kappa_+}} \anhb_{\omega,\vec{\ell}\,}^{\dagger}|\Psi\rangle .
\end{equation}
Clearly, the smoothness of the quantum state requires entanglement of modes across the horizon. This entanglement enables the mirror operator construction of the interior modes \cite{Papadodimas:2013jku,Papadodimas:2013wnh}. 

Using \eqref{actionet}, we can determine $\langle\Psi| \anhe_{\omega,\vec{\ell}\,} \anhet_{\omega,\vec{\ell}\,}|\Psi\rangle$,
\begin{equation}
\langle\Psi| \anhe_{\omega,\vec{\ell}\,} \anhet_{\omega,\vec{\ell}\,}|\Psi\rangle = e^{-{\pi \omega\over \kappa_+}}\langle\Psi| \anhc_{\omega,\vec{\ell}\,} \anhb_{\omega,\vec{\ell}\,}^{\dagger}|\Psi\rangle.
\end{equation}
The simplest way to compute the above expression is to express the near horizon modes in terms of global modes \eqref{localglobalmodesrelation}, use \eqref{bogexterior} to express $c_{\omega,\vec{\ell}\,}$ in terms of $a_{\omega,\vec{\ell}\,}$ and $b_{\omega,\vec{\ell}\,}$, and finally use the two-point functions of global modes in the exterior \eqref{2pointglobal} and \eqref{abdaggerglobal}. 
\begin{equation}\label{eet}
\begin{split}
\langle\Psi| \anhe_{\omega,\vec{\ell}\,} \anhet_{\omega,\vec{\ell}\,}|\Psi\rangle &= -({\kappa_{+}x_0\over\zeta_+})^{-2i\omega\over\kappa_+}{\cal X}(\omega),\\
{\cal X}(\omega) &=  { e^{-{\pi\omega\over\kappa_+}} \over \boga } \left[{1\over\bogb+\bogbs}\left({1\over1-e^{-{2\pi\omega\over\kappa_c}}}-{1\over 1-e^{-{2\pi\omega\over\kappa_+}}}\right) - {\bogbs\over 1-e^{-{2\pi\omega\over\kappa_+}}}\right].
\end{split}
\end{equation}
As explained in previous section, the two-point functions of the global modes is constrained by two-point functions of the local modes. The coefficient of the delta function in frequency is completely fixed. Using \eqref{eedagger}, \eqref{etetdagger} and \eqref{eet}, we get the following results.
\begin{equation}\label{2ptfnglobalint}
\begin{split}
\langle\Psi|e_{\omega,\vec{\ell}\,}e^{\dagger}_{\omega',\vec{\ell}\,}|\Psi\rangle &= {1\over 1- e^{- {2\pi \omega \over \kappa_+}}}\delta(\omega-\omega') +\cdots, \\
\langle\Psi|\tilde{e}_{\omega,\vec{\ell}\,}\tilde{e}^{\dagger}_{\omega',\vec{\ell}\,}|\Psi\rangle &= {1\over 1- e^{- {2\pi \omega \over \kappa_+}}}\delta(\omega-\omega')+\cdots ,\\
\langle\Psi|e_{\omega,\vec{\ell}\,}\tilde{e}_{\omega',\vec{\ell}\,}|\Psi\rangle &= {\cal X}(\omega)\delta(\omega-\omega')+\cdots ,
\end{split}
\end{equation}
where the ellipsis denote terms regular at $\omega=\omega'$.

\subsubsection{Two-point function at inner horizon}
By assuming smoothness of future and past cosmological horizon, and future outer horizon, we have managed to fix \emph{all} two point functions of the near horizon modes. Now, we compute the two-point function of local modes near the inner horizon, $\anhf_{\omega,\vec{l}\,}$. To compute $\langle\Psi|\anhf_{\omega,\vec{\ell}\,}\anhf_{\omega,\vec{\ell}\,}^{\dagger}|\Psi\rangle$, we use the relation between local and global modes \eqref{localglobalmodesrelationinterior}, the Bogoliubov transformation \eqref{boginterior}, and the two-point functions of global modes in the interior \eqref{2ptfnglobalint}. This simple computation gives the following result.
\begin{equation}
\begin{split}
\langle\Psi| \anhf_{\omega,\vec{\ell}\,}\anhf_{\omega,\vec{\ell}\,}^{\dagger}|\Psi\rangle = {|\bogainside|^2+|\bogbinside|^2 e^{-{2\pi\omega\over\kappa_+}} \over 1- e^{-{2\pi\omega\over\kappa_+}}} + (\bogainside\bogbinside {\cal X}(\omega) + \bogasinside \bogbsinside {\cal X}^*(\omega)) ,
\end{split}
\end{equation}
where, recall that, ${\cal X}$ is defined in \eqref{eet}. We define the fractional difference,
\begin{equation}\label{fractionaldiff}
\eta_{\omega,\vec{\ell}\,} =\left (1-e^{-{2\pi\omega\over\kappa_-}}\right)\langle\Psi| \anhf_{\omega,\vec{\ell}\,}\anhf^{\dagger}_{\omega,\vec{\ell}\,}|\Psi\rangle - 1.
\end{equation}
A non-zero fractional difference, for any frequency and angular momentum, would imply that the quantum state is \emph{not} smooth at the inner horizon, and hence, the strong cosmic censorship is not violated. To compute the fractional difference $\eta$, we only need to compute the Bogoliubov coefficients $\boga,\bogb,\bogainside,\bogbinside$. We solve the radial wave equation in the exterior and interior numerically and find that the inner horizon is not smooth, see section \ref{s_numerics} for details.  

We point out that this test is much simpler to implement than computing the full renormalized quantum stress tensor. To compute the stress tensor, we need to sum the contribution of all frequencies and angular momenta, and deal with ambiguities in the renormalization procedure. In contrast, to compute $\eta$, we only need to solve the wave equation for a particular frequency and angular momentum.



\section{Details of numerical computations and results \label{s_numerics}}
\subsection{Numerical algorithm and error estimates}
We solve the differential equation \eqref{radialwave} numerically and determine the Bogoliubov coefficients, $\boga$, $\bogb$, $\bogainside$ and $\bogbinside$. In the exterior, we solve from $r_+ + \epsilon$ to $r_c-\epsilon$, with plane wave initial condition near outer horizon, $F^N(r_*\rightarrow-\infty) \sim {e^{- i \omega r_*} \over  \sqrt{2\omega}}$. As this is a valid approximation close to the horizon, we choose $\epsilon=10^{-8}$. We work with black holes of size, $ r_+\sim {\cal O}(1)$, hence, $\epsilon\ll r_+$. The plane wave solutions are valid approximations near the cosmological horizon as well. To determine the Bogoliubov coefficients, we solve the following set of linear equations,
\begin{equation}
\begin{split}
\left.F^N(r_*) \right\vert_{r=r_c-\epsilon} & = \boga \left.{e^{-i\omega r_*}\over \sqrt{2\omega}}\right\vert_{r=r_c-\epsilon} + \bogb\left. {e^{i\omega r_*}\over \sqrt{2\omega}} \right\vert_{r=r_c-\epsilon},\\
\left.\dfrac{d}{dr_*}{F}^N(r_*)\right\vert_{r=r_c-\epsilon} & = -i\boga \left.{\sqrt{\omega} e^{-i\omega r_*}\over \sqrt{2}}\right\vert_{r=r_c-\epsilon} + i\bogb \left.{\sqrt{\omega} e^{i\omega r_*}\over \sqrt{2}}\right\vert_{r=r_c-\epsilon}.
\end{split}
\end{equation}
The determination of Bogoliubov coefficients in the interior can be carried out in the same way. Once we have determined all the Bogoliubov coefficients, we can numerically evaluate the fractional difference $\eta$ defined in \eqref{fractionaldiff}.

\subsection*{Error estimate}
Before proceeding to the results, we also estimate the error in numerical determination of the fractional difference. There are two sources of error. First, the error associated with the plane wave approximation near horizon. Second, the error associated with the numerical solution of the wave equation.
\paragraph{Error associated to the plane wave approximation near horizons. \\}
We will first estimate the error associated with the plane wave approximation. To remove redundancy, we only describe error estimation just outside the outer horizon. The form of potential near the outer horizon is,
\begin{equation}\label{potentialnearhorizon}
\begin{split}
V(r_+ +\epsilon) & =   - \omega^2 +  {2 \alpha \kappa_+ }\epsilon + {\cal O}(\epsilon^2),\\
\alpha &= {1\over r^2_+} \left( \ell(\ell+d-2) +(d-1)r_+\kappa_+ \right).
\end{split}
\end{equation}
Suppose, we express the exact solution to the differential equation, \eqref{radialwave}, as follows.
\begin{equation}
\begin{split}
\psi(r_*) &= \lambda(r_*) {e^{- i \omega r_*} \over  \sqrt{2\omega}},\\
\lambda(r_*) &= 1 + \sum_{n>0} s_n \epsilon^n.
\end{split}
\end{equation}
Recall that $dr/dr_*=f(r)$, and near outer horizon, $f(r_++\epsilon) = f'(r_+) \epsilon + {\cal O}(\epsilon^2) = 2 \kappa_+ \epsilon + {\cal O}(\epsilon^2)$. Hence,
\begin{equation}
\begin{split}
{d\over dr_*}\lambda(r_*) &=  2 s_1 {\kappa_+\epsilon} + {\cal O}(\epsilon^2),\\
{d^2\over dr^2_*}\lambda(r_*) &= 4 s_1 {\kappa^2_+ \epsilon }+ {\cal O}(\epsilon^2).
\end{split}
\end{equation}
Substituting in the differential equation and using the approximation for potential, \eqref{potentialnearhorizon}, we get,
\begin{equation}
\begin{split}
\lambda(r_*) = 1 + {\alpha \epsilon \over {2(\kappa_+- i \omega)}} + {\cal O}(\epsilon^2).
\end{split}
\end{equation}
Hence, the error associated with the plane wave approximation is,
\begin{equation}
\begin{split}
\left|\delta \psi \right|\approx {1\over\sqrt{2\omega}}{\alpha\,\epsilon \over 2\left|\kappa_+- i \omega\right|}, \qquad
\left|\delta {d{\psi}\over d r_*}\right| \approx {1\over\sqrt{2\omega}}{\alpha\,\epsilon \left|2\kappa_+- i \omega\right|\over 2\left|\kappa_+- i \omega\right|} . 
\end{split}\end{equation}
Similarly, we can compute the error in estimation of the solutions near cosmological horizon and inner horizon.
\paragraph{Error associated with the numerical evaluation of solutions of the wave equation.\\}
When we evaluate $F^N$ near the cosmological horizon, we get some error due to the numerical procedure implemented to solve the differential equation. To estimate this error, we set $F^N(r_c-\epsilon)$ as the initial condition and solve the differential equation from cosmological horizon to the outer horizon. Using the output as the initial condition, we once again solve the differential equation from outer horizon to the cosmological horizon. The final result differs from $F^N(r_c-\epsilon)$, and the difference is an estimate of the error associated with numerical solution, $\delta F^N(r_*)$. 

Taking these two sources of error into account, we estimate the error in determination of the Bogoliubov coefficients, $\delta\boga$, $\delta\bogb$, $\delta\bogainside$ and $\delta\bogbinside$. Using this, we estimate the error in fractional difference, $\delta\eta_{\omega,\ell}$. In all the plots, we display error bars as well.

\subsection{Summary of results}
\begin{figure}[h!]
	\centering
	\begin{subfigure}{0.48\textwidth}
		\includegraphics[width=\textwidth]{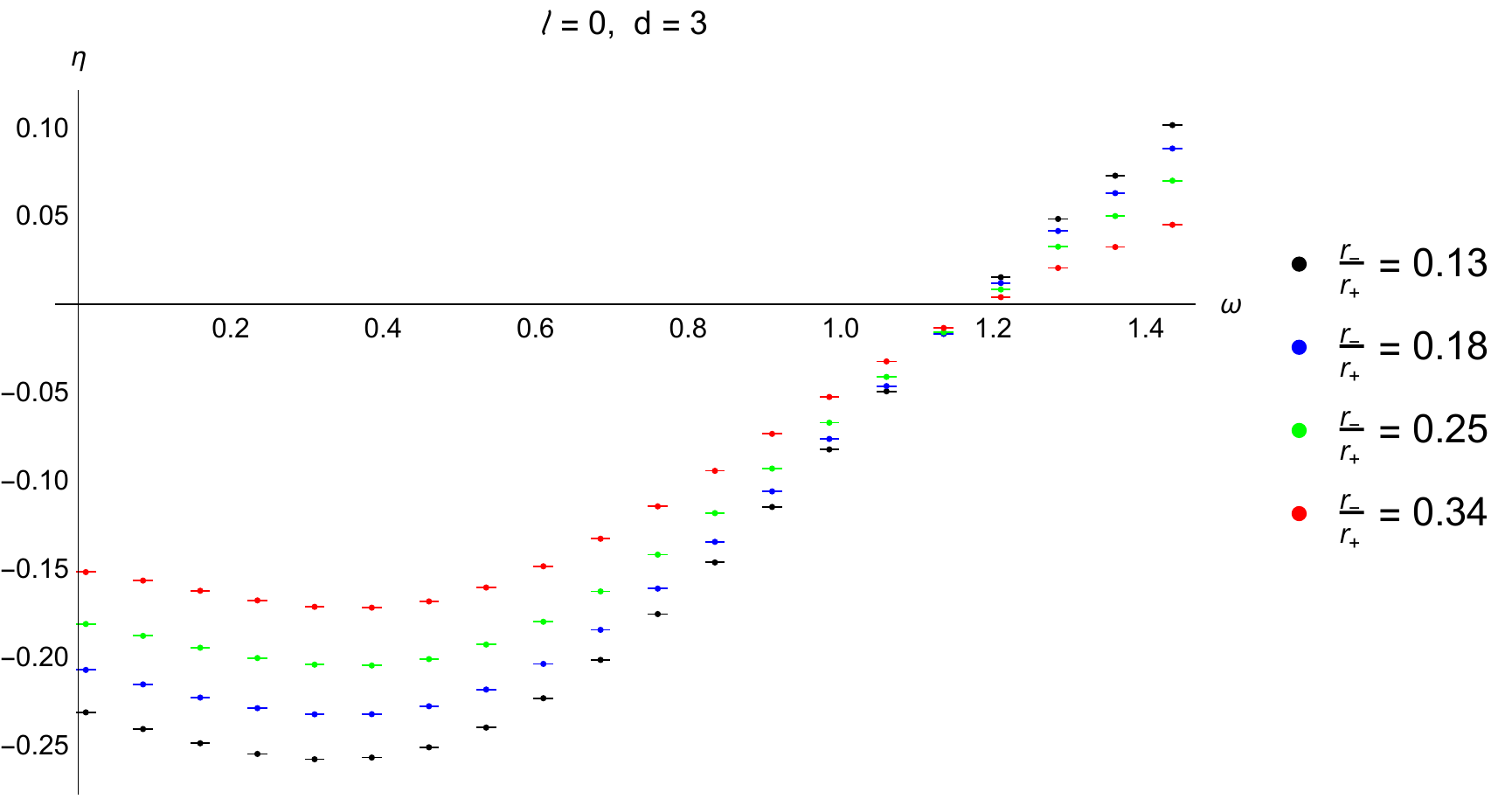}
	\end{subfigure}\hspace{8pt}
	\begin{subfigure}{0.48\textwidth}
		\includegraphics[width=\textwidth]{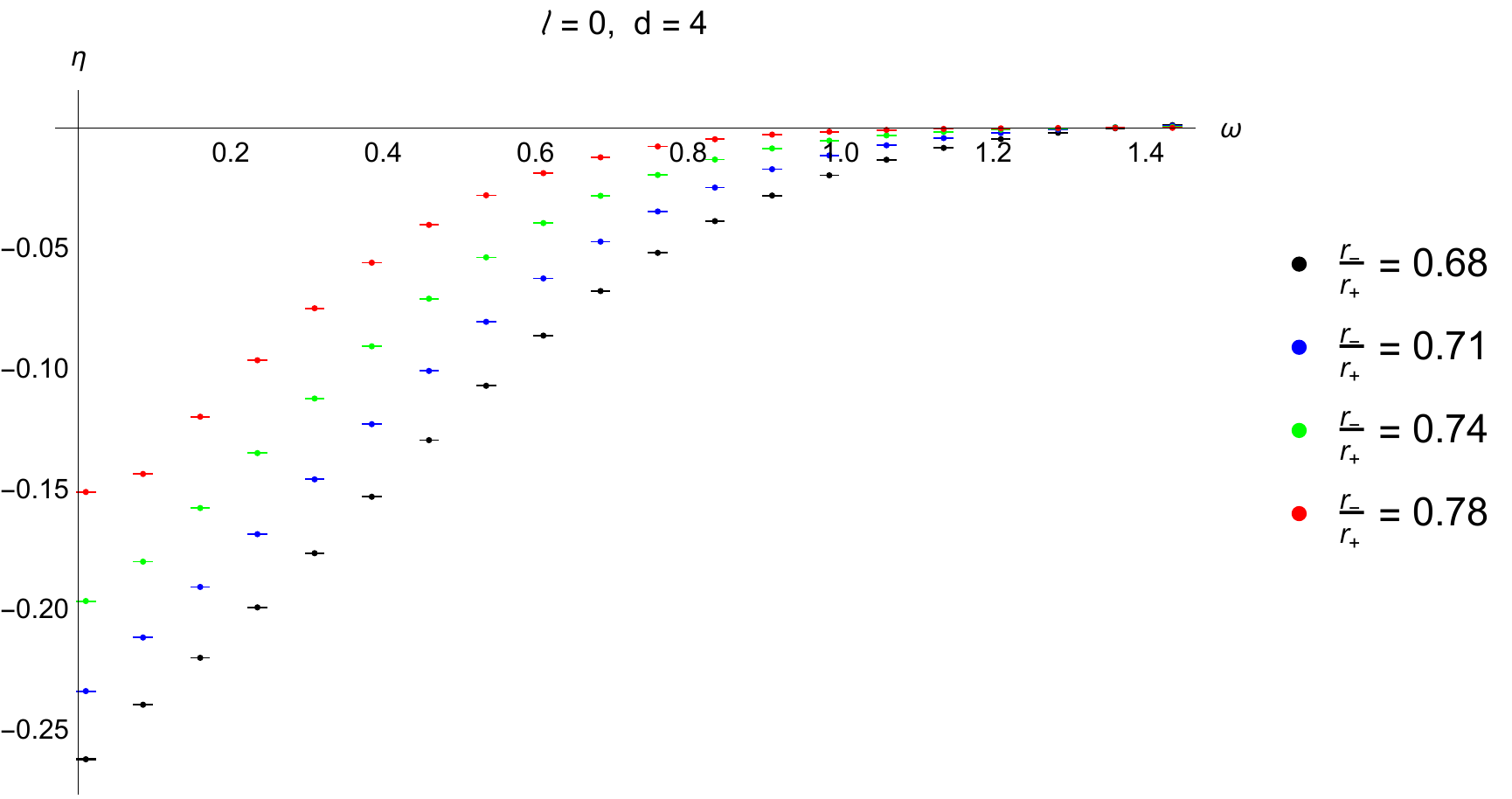}
	\end{subfigure}
	\vspace{10pt}
	\begin{subfigure}{0.48\textwidth}
		\includegraphics[width=\textwidth]{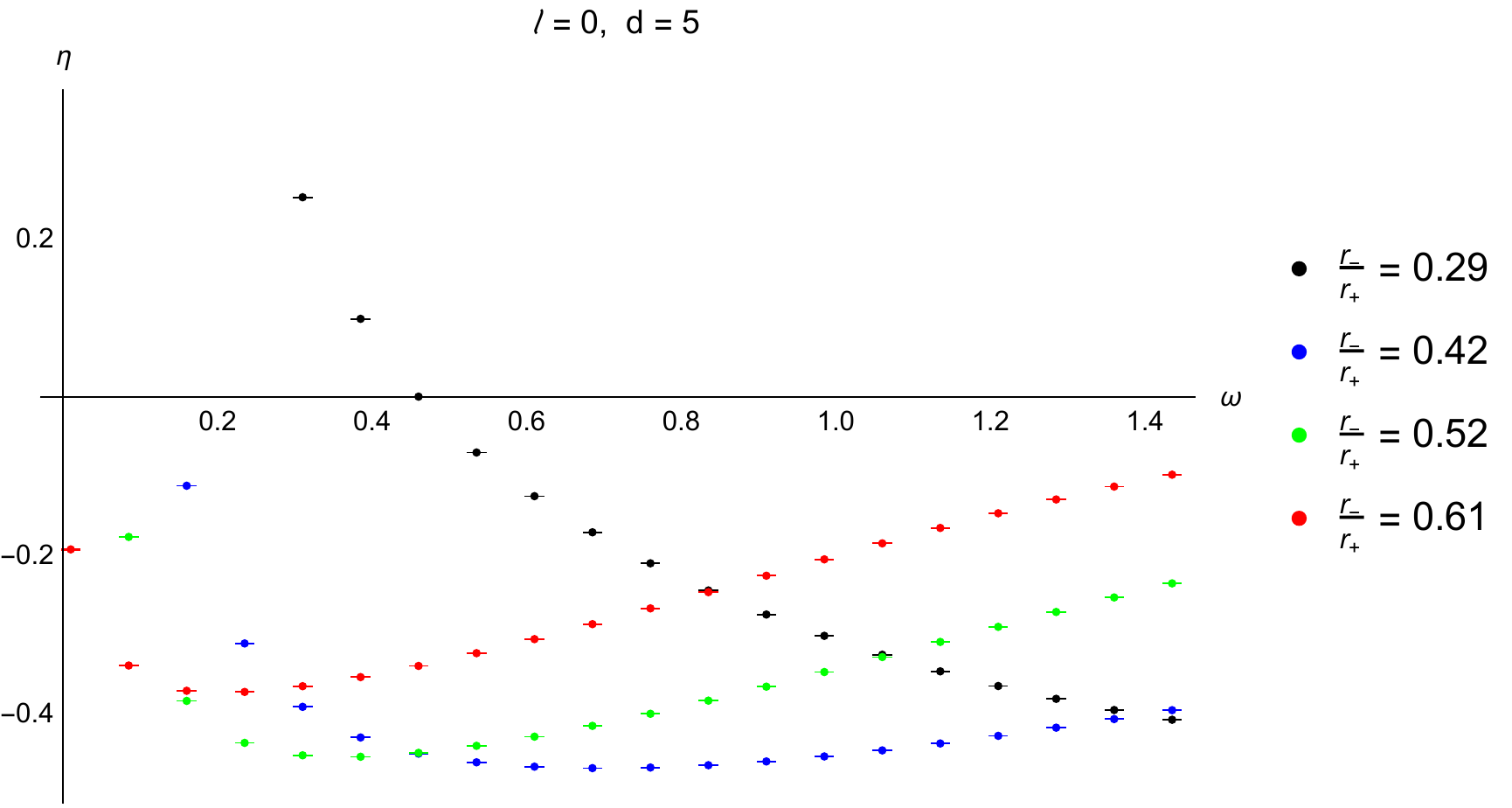}
	\end{subfigure}
	\caption{Plot of $\eta$ with $\omega$ for $\ell=0$ and $d=3,4,5$. We vary the ratio $r_-/r_+$. Since $\eta$ is non-zero for a range of frequencies, the inner horizon is unstable.}
	\label{fig_etavsw}
\end{figure}

\FloatBarrier
We run the simulations for various parameters in 3, 4 and 5 spatial dimensions. First, we plot the variation of fractional difference with frequency for $\ell=0$, $d=3,4,5$, and various values of $r_+$ and $r_-$, Fig. \ref{fig_etavsw}. Since the fractional difference is not zero for all frequencies, the local modes near the inner horizon are not thermally populated with temperature $2\pi\over\kappa_-$. As a consequence, the state $|\Psi\rangle$ is singular at the inner horizon. Hence, no quantum state in the Hilbert space is smooth at all three horizons.

\begin{figure}[h!]
	\centering
	\vspace{30pt}
	\begin{subfigure}{0.48\textwidth}
		\includegraphics[width=\textwidth]{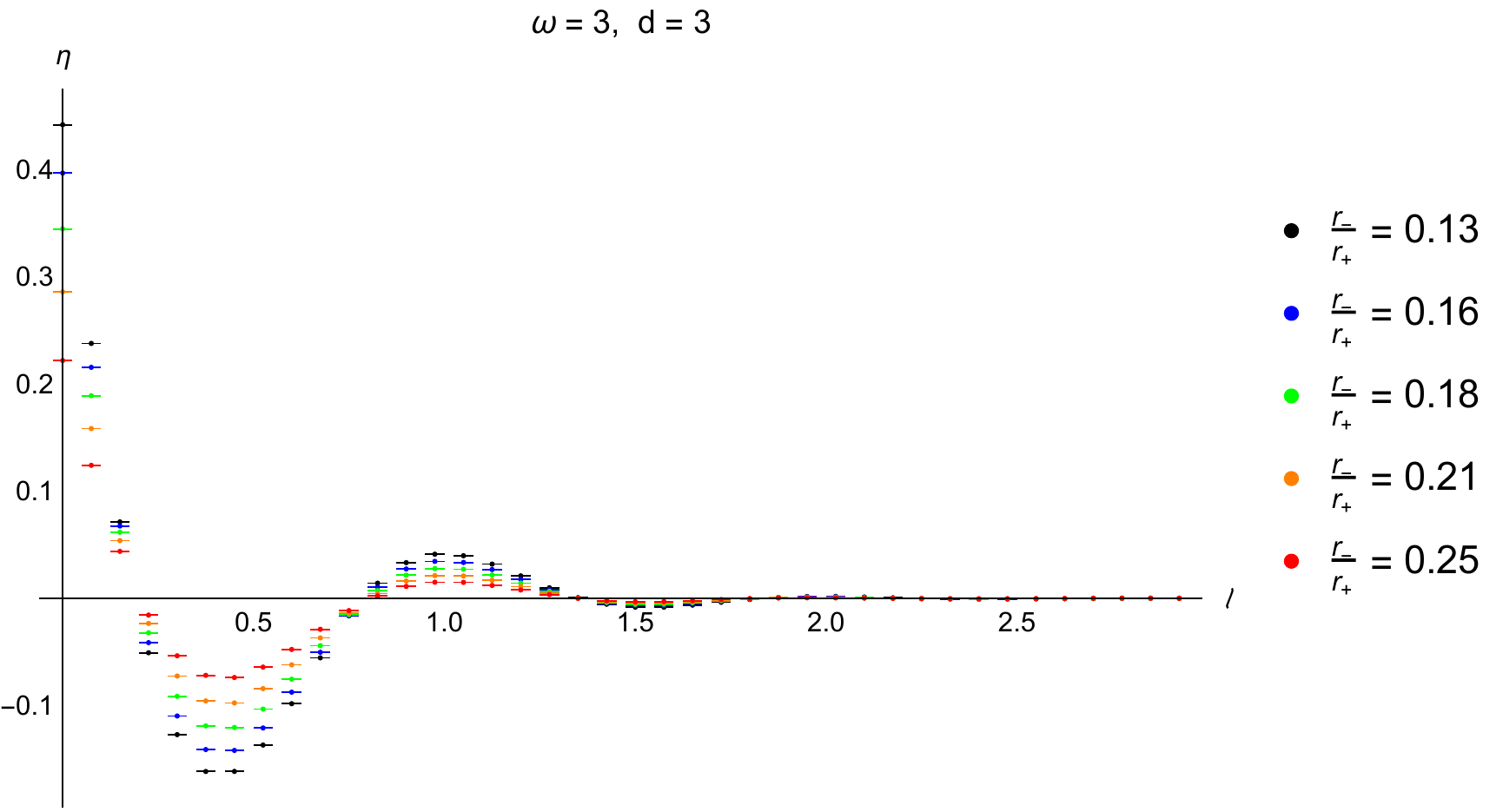}
	\end{subfigure}\hspace{8pt}
	\begin{subfigure}{0.48\textwidth}
		\includegraphics[width=\textwidth]{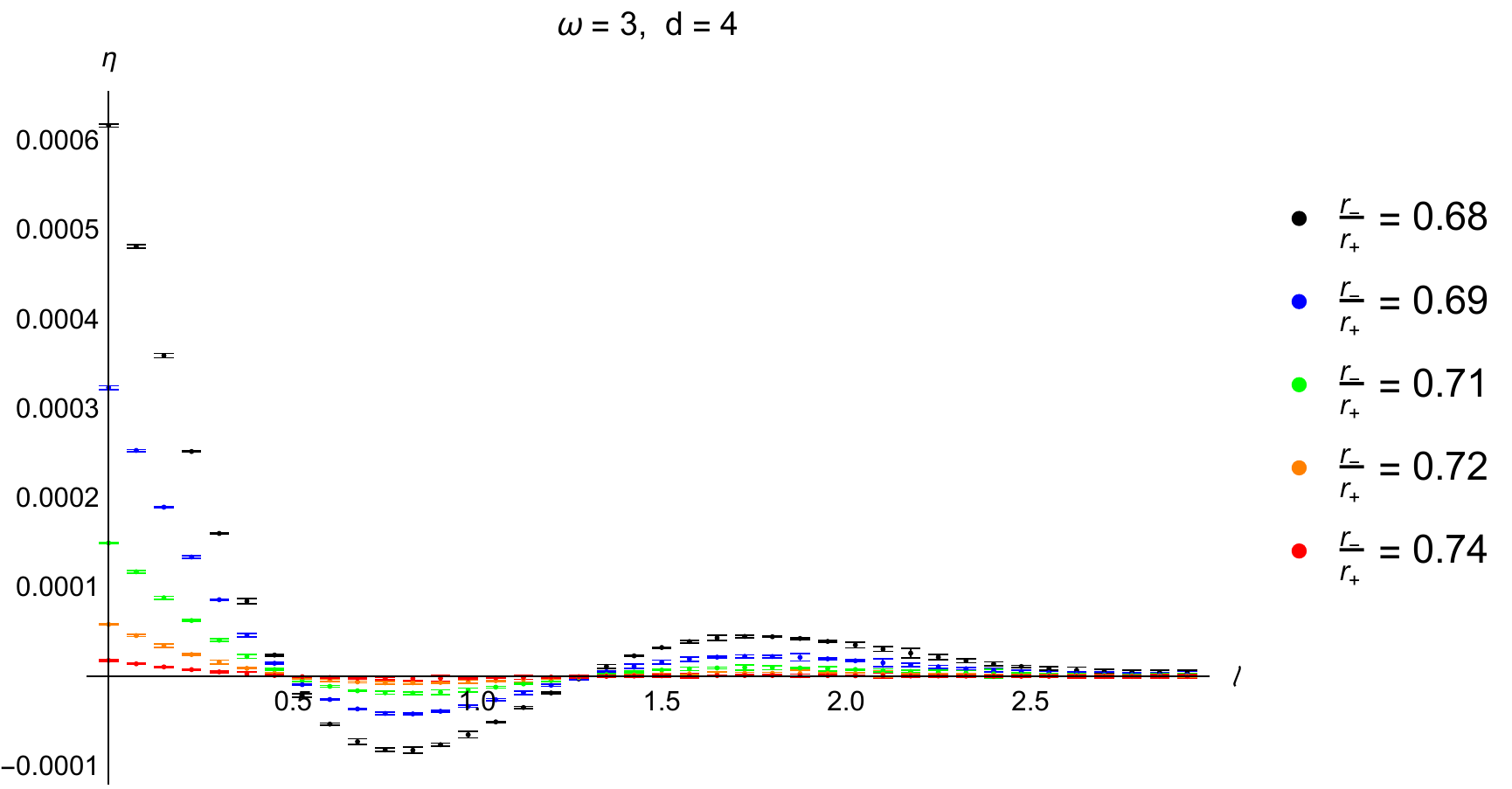}
	\end{subfigure}\vspace{10pt}
	\begin{subfigure}{0.48\textwidth}
		\includegraphics[width=\textwidth]{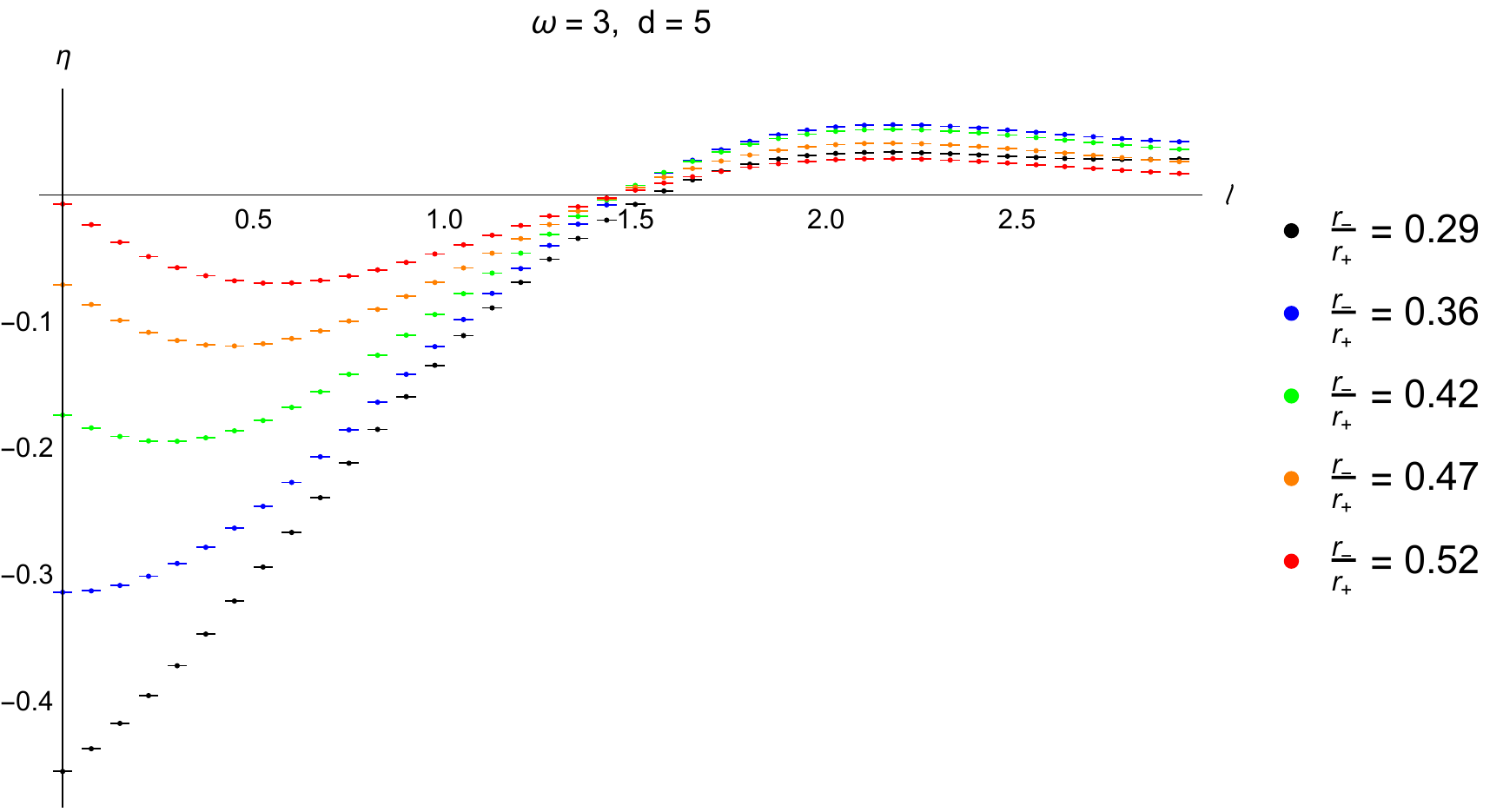}
	\end{subfigure}
	\caption{Plot of $\eta$ vs $\ell$ for $\omega=3$ and $d=3,4,5$. As expected, $\eta$ goes to zero for large $\ell$.}
	\label{fig_etavsl}
\end{figure}

One can analytically show that at large angular momentum, $\ell$, the fraction difference should go to zero, see Appendix of \cite{Papadodimas:2019msp}. This is related to the fact that coefficient of the leading order divergence in $\langle \phi^2 \rangle$, controlled by large-$\ell$, is zero \cite{Sela:2018xko}. In Fig. \ref{fig_etavsl}, we verify these expectations numerically. This provides a non-trivial consistency check for our numerical algorithm.

\begin{figure}[h!]
	\centering
	\begin{subfigure}{0.48\textwidth}
		\includegraphics[width=\textwidth]{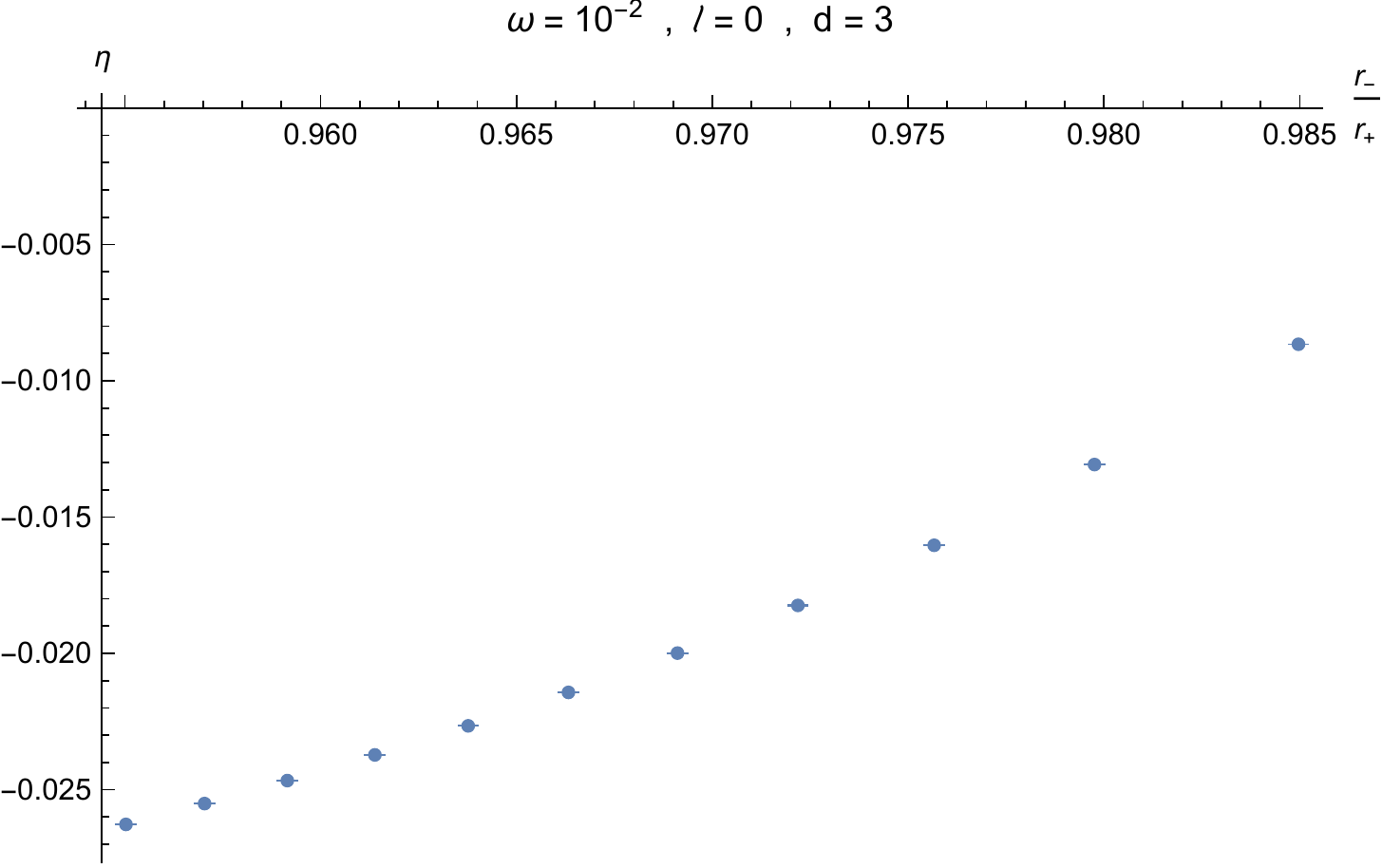}
	\end{subfigure}\hspace{8pt}
	\begin{subfigure}{0.48\textwidth}
		\includegraphics[width=\textwidth]{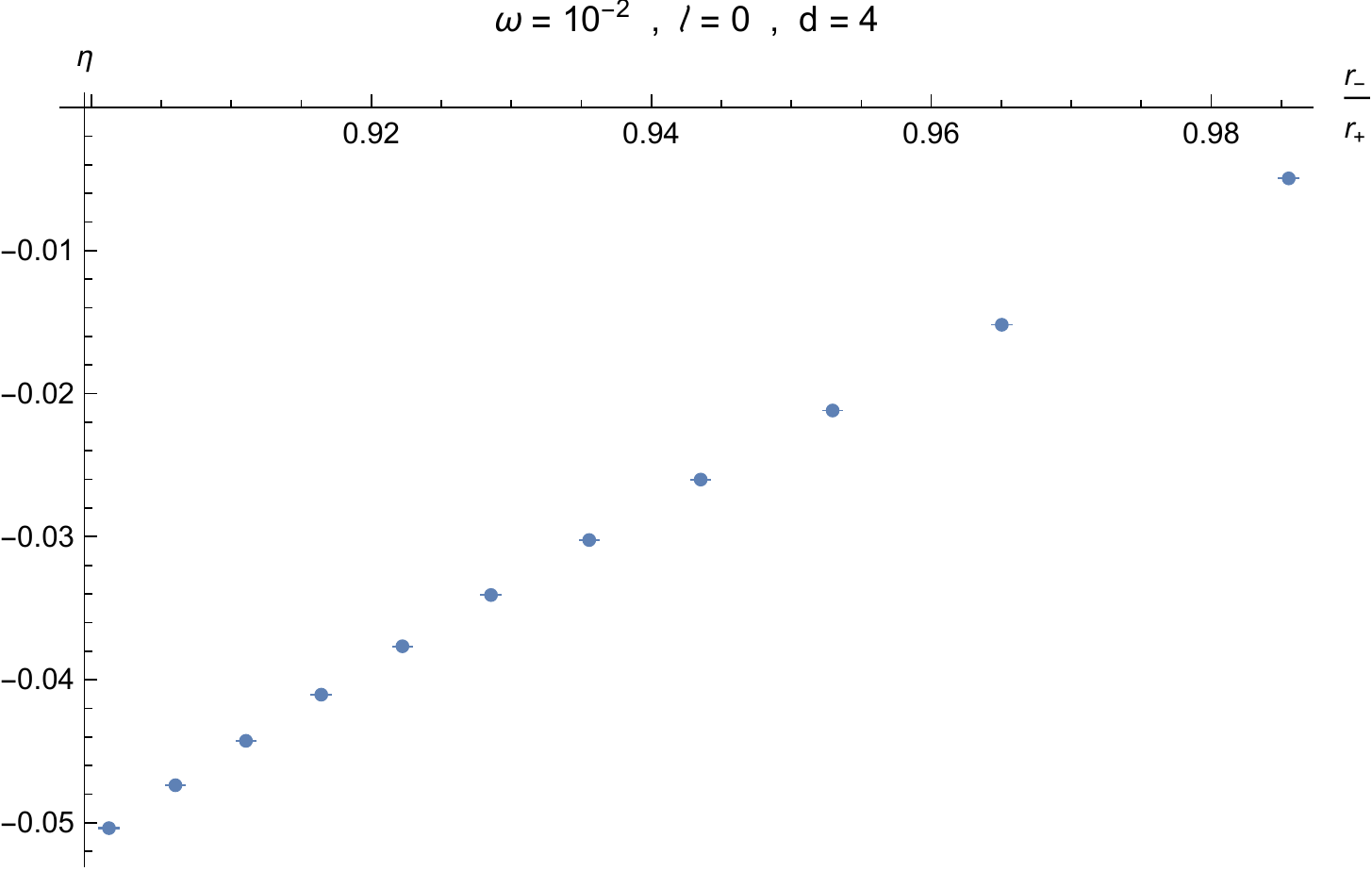}
	\end{subfigure}\vspace{10pt}
	\begin{subfigure}{0.48\textwidth}
		\includegraphics[width=\textwidth]{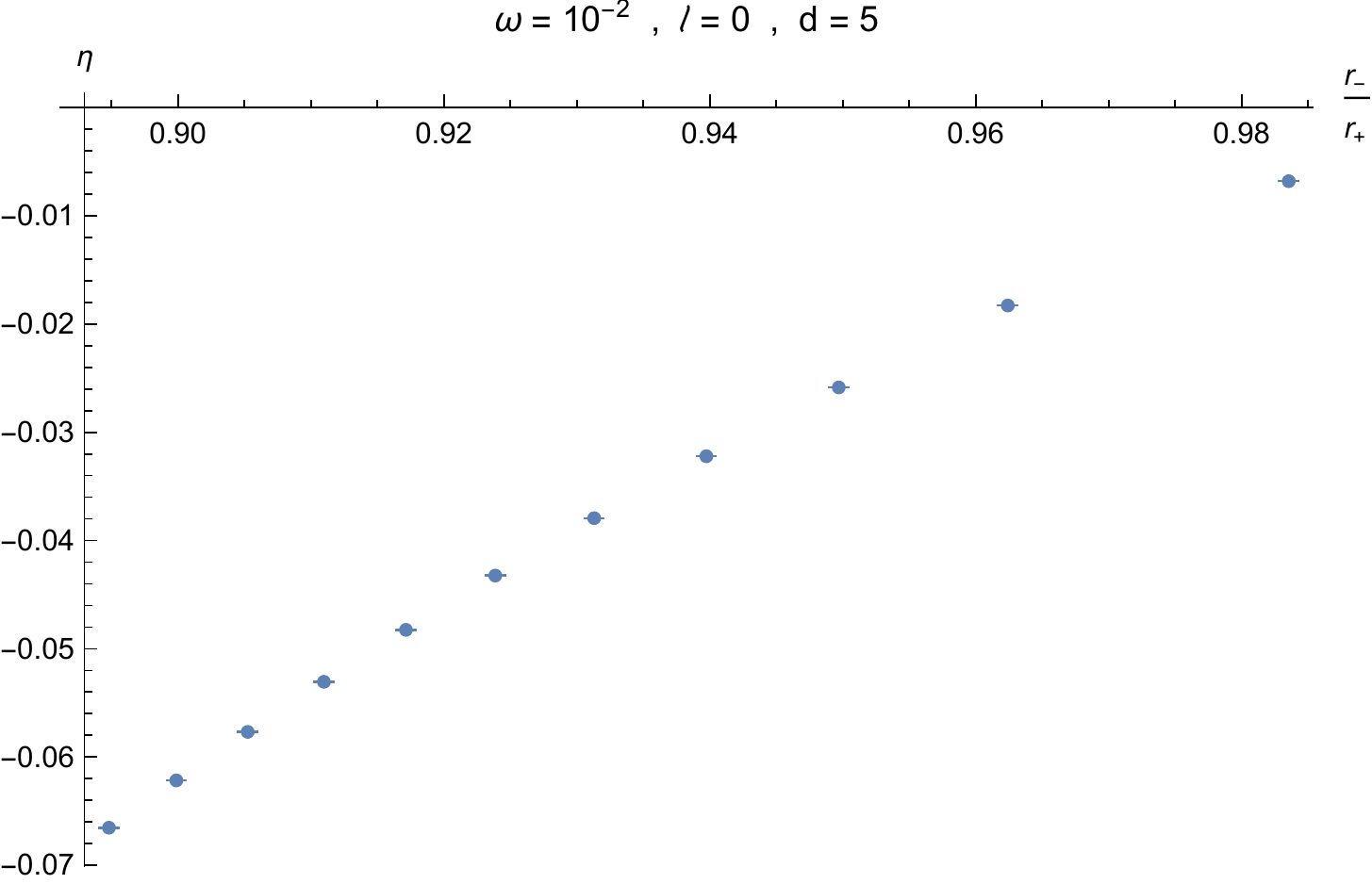}
	\end{subfigure}
	\caption{Plot of $\eta$ with $r_-\over r_+$ for $\omega=0.01$, $\ell=0$ and $d=3,4,5$. The fractional difference is non-zero even close to extremality.}
	\label{fig_etavsrmoverrp}
\end{figure}

The violations of strong cosmic censorship under classical perturbations has been demonstrated for RN-dS black holes close to extremality \cite{Cardoso:2017soq,Dias:2018ufh}. In Fig. \ref{fig_etavsrmoverrp}, we plot fractional difference for various values of the ratio of radius of inner horizon to the radius of outer horizon, $r_-/r_+$. We fix the frequency, $\omega=10^{-2}$, and angular momentum, $\ell=0$. We find that even close to extremality, the fractional difference is non-zero. In \cite{Cardoso:2017soq}, violations of strong cosmic censorship were observed for parameter $B$, close to 99\% of its extremal value. This corresponds to ${r_-\over r_+} \sim 0.75$. However, as is evident from Figure \ref{fig_etavsrmoverrp}, we probe black holes much closer to extremality, ${r_-\over r_+} > 0.9$, and still find that quantum fluctuations destabilize the inner horizon.

From these results it is clear that the modes near inner horizon are not correctly entangled, even close to extremality. As two point functions near inner horizon in state $|\Psi\rangle$ fails to reduce to the flat space limit when the insertions are coincident, the stress tensor at the inner horizon diverges. Hence, the inner horizon is rendered unstable by quantum fluctuations. Extension of spacetime beyond the inner horizon is ruled out and the strong cosmic censorship conjecture is restored in asymptotically de-Sitter Reissner-Nordstr\"om black holes.



\section{Conclusions}\label{s_conclusions}
In this paper, we addressed the question of validity of strong cosmic censorship conjecture in Reissner-Nordstr\"om black hole in de-Sitter space. Violations of the conjecture under classical perturbations was established in \cite{Cardoso:2017soq,Dias:2018ufh}. However, we argued that quantum fluctuations are sufficient to restore the conjecture. To achieve this, we studied the propagation of a quantum scalar field in the fixed RN-dS background and showed that the inner horizon is unstable.

The smoothness of a quantum state across a null surface requires a universal entanglement of ``near horizon'' modes across the surface \cite{Papadodimas:2019msp}. Before answering the question of smoothness of the inner horizon, it is important to study the existence of a quantum state that is smooth in the exterior region. Due to the presence of cosmological horizon in asymptotically de-Sitter spacetimes, which generically radiates at a different temperature than the event horizon, existence of such a smooth state is not obvious. However, in section \ref{s_exterior}, we demonstrated the existence of such a state for all spherically symmetric black holes in de-Sitter space, and in arbitrary dimensions.

To be precise, we demonstrated that the two point function of the scalar field insertions have the correct flat space limit near event horizon and cosmological horizon, \eqref{twoptfn}. This ensures that the coefficient of the leading order divergence in the stress tensor is zero. Even though this doesn't necessarily imply that the stress tensor is finite, it is strongly suggestive of the possibility, and demands further analysis.

Demanding smoothness of states across cosmological and outer horizon, completely fixes all two-point functions of the near horizon modes, even in the interior (see section \ref{s_interior}). In particular, this enables us to compute the two point function of the local modes defined near inner horizon. Smoothness of inner horizon necessarily requires the fractional difference $\eta_{\omega,\vec{\ell}\,}$, defined in \eqref{fractionaldiff}, to vanish. Note that, these are infinitely many constraints, one for each frequency and angular momentum. 

The evaluation of $\eta_{\omega,\vec{\ell}\,}$ only requires the determination of Bogoliubov coefficients, and can be obtained by solving the radial wave equation for modes with frequency $\omega$ and angular momentum $\vec{\ell}$. In section \ref{s_numerics}, we evaluated the Bogoliubov coefficients through a simple numerical computation. 

We found that the fractional difference is non-zero for various mass, charge and dimension of the black hole. We also explored black holes very close to extremality, and found the same result. The non-vanishing of fractional difference implies that the stress tensor at the inner horizon diverges. This instability is sufficiently strong to rule out violations of the strong cosmic censorship conjecture. The metric cannot be continued beyond the inner horizon, even as a weak solution. The spacetime must end in a singularity near the inner horizon.

If we restrict to the study of classical perturbation, then there is always a regime of parameters close to extremality for which determinism fails. To restore determinism under classical perturbations, it has been suggested that we should only allow \emph{rough} initial data, since scalar perturbation at Cauchy slice is generically less regular than the initial data \cite{Dafermos:2018tha,Dias:2018etb}. However, once we include quantum corrections, such an initial condition is no longer required. This demonstrates the importance of understanding the implications of quantum effects for the cosmic censorship conjecture.

Our results are in agreement with the results of \cite{Hollands:2019whz,Hollands:2020qpe}, where the coefficient of leading order divergence in the stress tensor was shown to be non-zero, numerically. However, the computation of the stress tensor requires mode summation and renormalization, and hence, is difficult to implement. On the other hand, our test only requires solving the wave equation for a particular mode, and therefore its implementation is much more convenient.

Finally, with slight modifications, our analysis can be easily generalized to non-spherically symmetric spacetimes, such as rotating black holes. We suspect that even for such black holes, the inner horizon would be destabilized by quantum effects.

\section*{Acknowledgments}
I am very grateful to Kyriakos Papadodimas and Suvrat Raju for collaboration in the early stages of this work, and various helpful discussions and comments. I would like to thank Chandramouli Chowdhury, Ben Freivogel, Victor Godet, Chandan Jana and Siddharth Prabhu for useful discussions. I would also like to thank Chethan Krishnan and Alok Laddha for comments on the draft of this manuscript. This work was partly completed at the International Centre for Theoretical Sciences, Bengaluru.



\bibliographystyle{JHEP}
\bibliography{references}
\end{document}